\documentclass[10pt,onecolumn, draftcls]{IEEEtran}

\usepackage{amsfonts,amsmath,amssymb,amsxtra}
\usepackage{subfigure,graphics,graphicx,epsfig,color}
\usepackage[ruled,vlined,linesnumbered]{algorithm2e}

\newcommand{\argmin}{\operatornamewithlimits{argmin}}

\newcommand{\sldots}{\mbox{{\small\ldots}}}

\newcommand{\mat}[1]{\mathbf{#1}}

\newtheorem{theorem}{Theorem}[section]

\newtheorem{lemma}[theorem]{Lemma}

\DeclareMathOperator{\Cov}{Cov}

\newcommand{\mapCoeff}{F}

\newcommand{\citeyear}{\cite}

\begin{document}

\title{Low-Complexity Coding and Source-Optimized Clustering
for Large-Scale  Sensor Networks}

\author{
GERHARD MAIERBACHER and JO\~AO BARROS
\\
Instituto de Telecomunica\c{c}\~{o}es,
\\
Department of Computer Science,
\\
Universidade do Porto,
\\
R. Campo Alegre 1021/1055, 4169-007 Porto, Portugal
\\
Email: \{gerhard, barros\}@dcc.fc.up.pt
}

\maketitle

\begin{abstract} 
We consider the distributed source coding problem in which correlated data
picked up by scattered sensors has to be encoded separately
and transmitted to a common receiver, subject to a rate-distortion constraint.
Although near-to-optimal solutions based on Turbo and LDPC codes exist for this problem,
in most cases the proposed techniques do not scale to networks of hundreds of
sensors. We present a scalable solution based on the following key elements:   
(a) distortion-optimized index assignments for low-complexity distributed quantization, 
(b) source-optimized hierarchical clustering based on the Kullback-Leibler distance and (c) sum-product decoding
on specific factor graphs exploiting the correlation of the data. 
\end{abstract}


\section{Introduction}
\label{s:intro}
\noindent 
In distributed sensing scenarios, where correlated data
has to be gathered by a large number of low-complexity,
power-restricted sensors, efficient source coding and data gathering
techniques are key towards reducing the required number of transmissions
and enabling extended network life-time.
Inspired by the seminal work of Slepian and
Wolf~\cite{slepian-wolf:correlated-sources}, characterizing the
fundamental limits of separate encoding of correlated sources,
several authors have contributed with distributed source coding solutions (see 
e.g.~\cite{DSCforSensorNetw} and references therein).
Focusing on scalar quantization, Flynn and Gray~\citeyear{FlynnG:87} provided 
one of the first practical approaches to construct distributed source codes 
for two continuous-valued sources. 
The basic idea behind this approach---which will also play
an important  role in our work---is to reuse the indices of a high-resolution
quantizer such that the overall end-to-end distortion after joint 
decoding is minimized.
Pradhan and Ramchandran presented in~\citeyear{sandeep-kannan:discus} 
a method called distributed source coding  
using syndromes (DISCUS), based on channel codes with good distance properties,
where the set of possible codewords is partitioned into co-sets and only
the co-set's syndrome and not the actual codeword is transmitted to the
decoder. This method, originally considered for an asymmetric scenario where 
information about one source is available as side information at the decoder,
was recently extended to the symmetric case~\cite{Pradhan:CosetCodes} where all sources are to be encoded and side information is not available at the decoder.
An alternative approach for the asymmetric scenario 
was provided by Zamir et al.~\citeyear{zamir00nested} and by Servetto in~\citeyear{Servetto:02b} where a constructive approach for Gaussian sources based on linear codes and nested 
lattices was presented.
Cardinal and Van Assche~\citeyear{CardinalAssche:EntropyQuant} as well as 
Rebollo-Monedero et al.~\citeyear{Rebollo-Monedero:QuantizersForDSC}
focused on the optimization of the quantization stage and proposed 
design algorithms for multiterminal quantizers.
A novel design concept for distributed source coding was presented in~\cite{maierbacherB:dio} where basic tools from fundamental number theory, specifically Diophantine analysis, are used to construct index assignments capable of exploiting statistical properties common to many important source models. 
Beyond these contributions, highly evolved iterative channel coding techniques 
such as low density parity check (LDPC) and turbo codes have been applied
to the distributed source coding problem~\cite{DSCforSensorNetw}, 
reaching the fundamental limits of Slepian and Wolf~\citeyear{slepian-wolf:correlated-sources}.

Despite these important contributions, very little is known on how to perform distributed compression in {\it large-scale} sensor networks (i.e. with hundreds of sensor nodes).
The main reason for this is that most approaches become infeasible
when the complexity of joint decoding or the complexity of 
a joint design of separate encoders is considered for a large number of correlated sources. 
Previous work towards this goal produced a scalable solution for the {\it decoding} side by running the sum-product algorithm on a carefully chosen factor graph approximation of the source correlation~\cite{bar05}.
In this paper, we present a scalable solution which includes the {\it encoding} side. The main idea is to reduce the number of quantization bits in a systematic way, exploiting correlation preserving clusters, which minimize the Kullback-Leibler Distance (KLD) between the given source statistics and a factor graph approximation.
Our main contributions are as follows:
\begin{itemize}
\item {\it Design of Low-Complexity Distributed Source Codes:} We propose a methodology to design quantizers for a very large number of sensors ($>100$) which exploits the spatial correlation between sensor measurements. 
Inspired by~\cite{FlynnG:87} we formulate a generalized index-reuse optimization algorithm which allows us to reduce the number of bits for data transmission by adding to our system a coarse quantization stage. 
\item {\it Source-Optimized Clustering:} We devise a hierarchical clustering algorithm that uses 
the joint probability density function (PDF) of the sensor measurements to partition the set of all sensors into clusters and prove that the complexity of quantizer design can be reduced significantly.
\item {\it Combination with Factor Graph Decoding:} We show how source-optimized clusters used for distributed source coding can be incorporated in a KLD optimized factor graph which, in turn, is used at the decoder to exploit source correlations in a  computationally tractable way.
\item {\it Simulation Results:} We show how our techniques can be applied to general sensor network scenarios as well as the so-called CEO problem~\cite{Berger:CEO} and provide numerical results for setups with $100$ encoders.
\end{itemize}
The rest of the paper is organized as follows. In Section~\ref{s:setup} we give a precise formulation of the problem setup and describe the underlying system model. 
In Section~\ref{s:index} we present a technique to optimize quantizers exploiting correlations in the source observations. 
Section \ref{s:sourceModel} describes our scalable solution based on source-optimized hierarchical clustering in sensor networks. 
The results of numerical experiments are discussed in Section~\ref{s:discussion}. The paper is concluded in Section~\ref{s:conclusion}.

\section{System Setup}
\label{s:setup}
\noindent
We start by introducing our notation. Random variables are always denoted by capital letters, e.g. $U$, where its realizations are denoted by the corresponding lowercase letters, e.g. $u$. 
Vectors are denoted by bold letters and, if not stated differently, assumed to be column vectors, e.g. $\mathbf{u}=(u_1,u_2,\sldots,u_N)^T$ and $\mathbf{U}=(U_1,U_2,\sldots,U_N)^T$.
The expression $\mathbf{0}_N = (0,0,\sldots,0)^T$ is the length-$N$ zero vector.
Matrices are denoted by bold capital letters, e.g. $\mat{A}$, where its determinant is referred to by the usage of vertical bars, e.g. $|\mat{A}|$.
The expression $\mat{I}_N$ is the $N \times N$ identity matrix.
It is always clear from the context, or stated explicitly, if a bold capital letter refers to a vector of random variables or to a matrix. 
Index sets are denoted by capital calligraphic letters, e.g. $\mathcal{N}$, unless otherwise noted, where the set's cardinality is referred to by the usage of vertical bars, e.g. $|\mathcal{N}|$. 
We follow the convention that variables indexed by a set denote a set of variables, e.g. if $\mathcal{N}=\{1,2,3\}$
then $u_{\mathcal{N}} = \{u_1,u_2,u_3\}$, and use the same concept
to define vectors of variables, e.g. $\mathbf{u}_{\mathcal{N}} = (u_1,u_2,u_3)^T$.
Furthermore, the entries of a vector are referred to by specifying its index within paretheses, e.g. $\mathbf{u}(0)$ refers to the first and $\mathbf{u}(N-1)$ to the last entry of the length-$N$ vector $\mathbf{u}$.

The covariance is defined by $Cov\{{\bf a},{\bf b}\}=E\{{\bf a}{\bf b}^T \}\!-\!E\{{\bf a}\}E\{{\bf b}\}^T$, where $E\{\cdot\}$ is the expectation
operator.

An $N$-dimensional random variable with realizations $\mathbf{u}=(u_1\,u_2,\cdots,u_N)^T \in \mathbb{R}^N$ is Gaussian distributed with mean $\pmb\mu= E\{\mathbf{u}\}$ and covariance matrix $\pmb\Sigma=\Cov\{\mathbf{u},\mathbf{u}\}$ when $p(\mathbf{u})$ is given by
\begin{equation}
  p(\mathbf{u})=
  \exp(-\frac{1}{2} ({\bf u}-{\pmb\mu})^T \pmb\Sigma^{-1}({\bf
u}-{\pmb\mu}))
  /((2 \pi)^N |\pmb\Sigma|)^{1/2}.
  \label{eqn:mvn_pdf}
\end{equation}
Such a PDF is simply denoted as $\mathcal{N}(\pmb\mu,\pmb\Sigma)$.
\begin{figure}[tb]
  \centering
  \includegraphics[width=8.2cm]{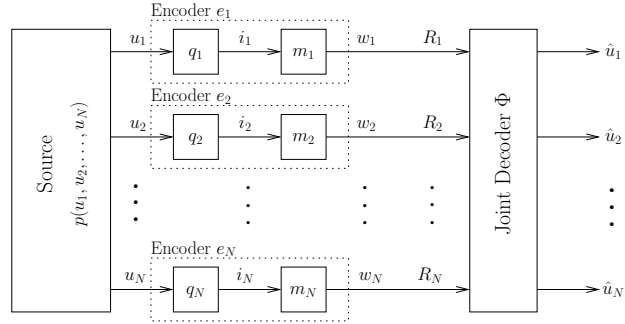}
\vspace{-0.2cm}
\caption{
System model. $N$ correlated sources are encoded independently and decoded jointly. At each encoder $e_n$, $n \in \{1,2,\sldots,N\}$, the observed source symbol $u_n$ is encoded onto the codeword
 $w_n$ and communicated to the joint decoder $\Phi$ at rate $R_n$. In the first stage of encoding, the discrete source index $i_n$ is obtained from $u_n$ by the scalar quantizer $q_n$ and, subsequently, $w_n$ is obtained by the index assignment $m_n$ such that $e_n = m_n \circ q_n$. After perfect transmission the joint decoder uses the vector of received codewords $\mathbf{w} = (w_1,w_2,\sldots,w_N)^T$ and its knowledge about the source statistics $p(u_1,u_2,\sldots,u_N)$ to jointly form the estimates $\mathbf{\hat{u}} = (\hat{u}_1,\hat{u}_2,\sldots,\hat{u}_N)^T$.}
%
  \label{fig:channelModel}
\end{figure}
\subsection{System Model} \label{s:systemm}
\noindent
We consider a setup of $N$ independently operating sensors. In this setup each sensor indexed by $n \in \mathcal{N}$, $\mathcal{N}=\{1,2,\cdots,N\}$,
observes a continuous-valued source sample $u_n(t)$ at time instant $t$.
For simplicity, only spatial correlations between measurements and not their temporal dependence is considered such that
the time index $t$ is dropped and only one time instant is considered.
The vector of source samples $\mathbf{u}=(u_1,u_2,\cdots,u_N)^T$, $\mathbf{u}\in \mathbb{R}^N$, at each time instant $t$
is assumed to be one realization of a $N$-dimensional Gaussian random variable distributed 
according to $\mathcal{N}({\bf 0}_N,\mat{R})$ with the vector of mean values $\pmb\mu = \mathbf{0}_N$ and the covariance matrix $\pmb\Sigma$ set equal to the correlation matrix
\[
\mat{R} = \left[\begin{array}{cccc}
       1 & \rho_{1,2} & \cdots & \rho_{1,N} \\
       \rho_{2,1} & 1 & \cdots & \rho_{2,N} \\
       \vdots     & \vdots  & \ddots & \vdots     \\
       \rho_{N,1} & \rho_{N,2} & \cdots & 1
       \end{array}\right],
\]
such that the individual source samples $u_n$, $n \in \mathcal{N}$, have zero mean $E\{u_n\}=0$, unit variance $Cov\{u_n,u_n\}=1$ and are correlated with $u_m$, $m \neq n$, $m \in \mathcal{N}$, according to the correlation coefficient $\rho_{n,m} = Cov\{u_n,u_m\}$.
Gaussian models for capturing the spatial correlation between sensors at
different locations are discussed in~\cite{ScaglioneS:02}
and models for the correlation coefficients of physical processes unfolding in
a field  can be found in~\cite{dietrich97fast}.

\begin{figure}[b]
\centering
\includegraphics[width=6.5cm]{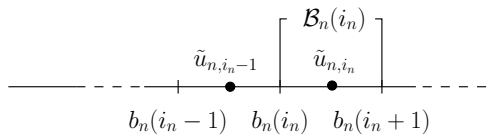}
\vspace{-0.4cm}
\caption{Scalar quantization. The source samples $u_n \in \mathbb{R}$ are mapped onto the index $i_n \in \mathcal{I}_n$ if they fall into the quantization region $\mathcal{B}_n(i_n)$ such that $b_n(i_n) < u_n \leq b_n(i_n+1)$.
Those samples, i.e. all $u_n \in \mathcal{B}_n(i_n)$, are then represented by their reconstruction level $\tilde{u}_{n,i_n} \in \mathcal{\tilde{U}}_n$.}
\label{fig:quant_enc}
\end{figure}

We assume that the sensors are low-complexity devices consisting only of
a scalar quantizer followed by an index assignment stage, see Figure~\ref{fig:channelModel}.
Specifically, we consider the following encoding procedure for each sensor $n \in \mathcal{N}$: 

In the first step, the observed source samples $u_n \in \mathbb{R}$ are mapped onto quantization indices $i_n \in \mathcal{I}_n$, $\mathcal{I}_n = \{0,1,\ldots,|\mathcal{I}_n|-1\}$, by the {\it quantization function} $q_n: ~~~ \mathbb{R} \rightarrow \mathcal{I}_n$ such that $i_n = q_n(u_n)$.
During quantization, an input value $u_n$ is mapped onto the index $i_n$ if it falls into the interval
$\mathcal{B}_n(i_n) \subseteq \mathbb{R}$ between the decision levels $b_n(i_n)$ and $b_n(i_n+1)$
such that $b_n(i_n) < u_n \leq b_n(i_n+1)$, see Figure~\ref{fig:quant_enc}.
The obtained quantization index $i_n$ is then associated with the reconstruction level $\tilde{u}_{n,i_n} \in \mathcal{\tilde{U}}_n$, $\mathcal{\tilde{U}}_n = \{\tilde{u}_{n,0}, \tilde{u}_{n,1}, \sldots, \tilde{u}_{n,|\mathcal{I}_n|-1}\}$, representing all source samples $u_n$ falling into the quantization region $\mathcal B_n(i_n)$.
We consider PDF optimized quantizers such that the mean squared error~(MSE) $E\{||U_n-\tilde{U}_n||^2\}=E\{(U_n-\tilde{U}_n)^2\}=\int_{u_n = -\infty}^{\infty}(u_n-\tilde{u}_{n,q_n(u_n)})^2\cdot~p(u_n)~du_n$ within the observations
is minimized, see e.g.~\cite{jayant-noll:coding-book}), which implies that the reconstruction levels $\tilde{u}_{n,i_n}$ are chosen to be the centroid (conditional expected value) of the quantization region $\mathcal B_n(i_n)$, i.e.
$\tilde{u}_{n,i_n} = E\{U_n|i_n\}$ for all $i_n \in \mathcal{I}_n$. 

In the second step of encoding, the obtained quantization index $i_n \in \mathcal{I}_n$ is mapped onto the codeword $w_n \in  \mathcal{W}_n$, $\mathcal{W}_n\!=\!\{0,1,\ldots,|\mathcal{W}_n|-1\}$, by the {\it mapping function}, also called the {\it index assignment}, $m_n: ~~~ \mathcal{I}_n \rightarrow \mathcal{W}_n$ such that $w_n = m_n(i_n)$.
We 
define the mapping function to be surjective, i.e. for any $w_n \in \mathcal{W}_n$ there exists at least one $i_n \in \mathcal{I}_n$ such that $w_n = m_n(i_n)$, for $n = 1,2,\sldots,N$. This property shall be important later on.

In summary, the encoder of each sensor operates in a sequential way and the overall {\it encoding function} 
can be expressed as $e_n = m_n \circ q_n$ such that $w_n= e_n(u_n) = m_n(q_n(u_n))$. 
The {\it data rate} at which the codewords $w_n$ are transmitted to the decoder is defined as $R_n = \lceil \log_2(|\mathcal{W}_n|) \rceil$ [bit].
%
%

Assuming data transmission over an array of $N$ ideal channels,
the decoder uses vector of codewords $\mathbf{w} = (w_1,w_2,\sldots,w_N)^T \in \mathcal{W}$, $\mathcal{W} = \prod_{n=1}^N \mathcal{W}_n$,
and available knowledge of the source correlation $\mat{R}$
to form the estimate $\mathbf{\hat{u}} = (\hat{u}_1, \hat{u}_2,\sldots,\hat{u}_N)^T$, $\mathbf{\hat{u}} \in \mathbb{R}^N$, of the originally observed source samples $\mathbf{u} \in \mathbb{R}^N$. The {\it decoding function} is defined as $\Phi: ~~~ \mathcal{W} \rightarrow \mathbb{R}^N$ such that $\mathbf{\hat{u}} = \Phi(\mathbf{w})$.
Assuming that the MSE $E\{||\mathbf{\hat{U}}\!-\!\mathbf{U}||^2\}$ between the estimates $\mathbf{\hat{U}} = (\hat{U}_1,\hat{U}_2,\sldots,\hat{U}_N)^T$ and source samples $\mathbf{U} = (U_1,U_2,\sldots,U_N)^T$ is the fidelity criterion to be minimized by the decoder, we observe that
\begin{equation}
E\{||\mathbf{\hat{U}}\!-\!\mathbf{U}||^2\} = E \{ (\mathbf{\hat{U}} - \mathbf{U})^T \cdot (\mathbf{\hat{U}} - \mathbf{U})\} = E\{\sum_{n=1}^N (\hat{U}_n - U_n)^2\} = \sum_{n=1}^N E\{(\hat{U}_n - U_n)^2\},
\end{equation}
which shows us that $E\{||\mathbf{\hat{U}}\!-\!\mathbf{U}||^2\}$ can be minimized globally by local minimization of the terms $E\{(\hat{U}_n - U_n)^2\}$ for $n=1,2,\sldots,N$.
The optimal estimate $\hat{u}_n(\mathbf{w})$ for a given codeword vector $\mathbf{w}$, i.e. such that $E\{(\hat{U}_n - U_n)^2\}$ is minimized globally, can be obtained by conditional mean estimation (CME), see e.g.~\cite{poor:detection+estimation-book}, such that
\begin{eqnarray} \label{eq:estimatesOpt}
\hat{u}_n(\mathbf{w}) = E\{U_n|\mathbf{w}\} \overset{\mathrm{(a)}}{=} \sum_{i_n=0}^{|\mathcal{I}_n|-1} E\{U_n|i_n\} \cdot p(i_n|\mathbf{w}) \overset{\mathrm{(b)}}{=} \sum_{i_n=0}^{|\mathcal{I}_n|-1} \tilde{u}_{n,i_n} \cdot p(i_n|\mathbf{w})
\end{eqnarray}
where equality~(a), as derived in Appendix~\ref{a:optDec}, allows us to express the estimate $\hat{u}_n(\mathbf{w})$ as a function of $E\{U_n|i_n\}$ and, thus, as a function of the reconstruction levels assuming that $\tilde{u}_{n,i_n} = E\{U_n|i_n\}$ as considered in~(b).

The required posterior probabilities $p(i_n=l|\mathbf{w})$ can be derived by
\begin{equation}
\label{eq:prob}
p(i_n=l|\mathbf{w}) \overset{\mathrm{(a)}}{=} \gamma \cdot p(i_n=l,\mathbf{w}) \overset{\mathrm{(b)}}{=} \gamma \cdot \sum_{\forall \mathbf{i} \in \mathcal{I}: i_n = l} p(\mathbf{w},\mathbf{i}), \\
\end{equation}
where the Bayes rule was applied in~(a) using the constant $\gamma\!=\!1/p(\mathbf{w})$ for normalizing the sum over all probabilities to one and in~(b) we calculate $p(i_n=l,\mathbf{w})$ from $p(\mathbf{w},\mathbf{i})$ by marginalizing over all possible realizations of $\mathbf{i} = (i_1,i_2,\sldots,i_N)^T$, $\mathbf{i} \in \mathcal{I}$, $\mathcal{I} = \prod_{\forall n \in \mathcal{N}} \mathcal{I}_n$. 
It is possible to express $p(\mathbf{w},\mathbf{i})$ in terms of the probability $p(\mathbf{i})$ known apriori from the source statistics and the transition probabilities $p(w_n|i_n)$ known from the index assignments $m_n$ for $n = 1,2,\sldots,N$ such that
\begin{equation}
\label{eq:prob2}
p(\mathbf{w},\mathbf{i}) \overset{\mathrm{(a)}}{=} p(\mathbf{w}|\mathbf{i}) \cdot p(\mathbf{i}) \overset{\mathrm{(b)}}{=} p(\mathbf{i}) \cdot \prod_{\forall n \in \mathcal{N}} p(w_n|i_n),
\end{equation} 
where the Bayes rule was applied in~(a) and~(b) takes into account that the index assignment operation performed at each encoder is independent from the other encoders. The probability mass function (PMF) $p(\mathbf{i})$ of the index vectors $\mathbf{i}$ can be obtained by numerically integrating the source PDF $p(\mathbf{u})$ over the quantization region defined by $\mathcal{B}_n(i_n)$ for all encoders $n = 1,2,\sldots,N$. Alternatively, one can resort to Monte Carlo simulation or approximate $p(\mathbf{i})$ by other means. 
Considering implementation issues it is worth pointing out that the transition probabilities $p(w_n|i_n)$ are either zero or unity since the mapping $m_n$ from the indices $i_n$ to the codewords $w_n$ is a function (i.e. knowledge of the index $i_n$ implies knowledge of the codeword $w_n$). Thus, the product of the transition probabilities in~(\ref{eq:prob2}) is also zero or unity, a fact, which can be exploited for an efficient implementation of the marginalization as shown in Appendix~\ref{a:marg}. 

The complexity of optimal decoding is analyzed in Appendix~\ref{a:implOptDec} and using the derived result we are able to state that the computational complexity of calculating all estimates according to~(\ref{eq:estimatesOpt}) is of $\mathcal{O}(N\mapCoeff^N)$ where $\mapCoeff \leq L-K+1$ is a system specific parameter depending on the characteristics of the mapping functions.\footnote{It is worth pointing out that decoding according to~(\ref{eq:estimatesOpt}) has only to be performed, in principle, only once for each realization $\mathbf{w} \in \mathcal{W}$ and that the calculated estimate could be then stored in the form of a decoding table for $n= 1,2,\sldots,N$. Thus, the decoding operation would reduce to a mere table look-up, i.e. decoding of $N$ sources would be of computational complexity $\mathcal{O}(N)$. However, such a decoding table itself has a space complexity of $\mathcal{O}(NK^N)$ and the computational complexity for creating it would be $NK^N$ times the complexity of decoding a single source, i.e. it would be of $\mathcal{O}(NK^N \mapCoeff^N)$. Therefore, the concept of using a decoding table shall be discarded throughout this work.}

\subsection{Our Goals} \label{s:ps}
\noindent
Under the system model described above, our first goal is to find distributed source coding algorithms that, by joint design of the index assignments, offers a suitable solution for large numbers of encoders.
Inspired by the work in \cite{FlynnG:87}, we formulate a generalized index-reuse algorithm to construct, for subsets of encoders, distortion-optimized index assignments suitable for distributed source coding.

Since optimal decoding according to~(\ref{eq:estimatesOpt}) is not feasible for large number of sources, part of this work shall be devoted to sub-optimal, yet feasible, decoders based on the principles presented in~\cite{bar05}.

We shall show that source-optimized clustering algorithms can be a key enabler towards the goal of obtaining both a scalable encoding and decoding solution feasible for large-scale sensor networks.

\section{Index Assignment Design}
\label{s:index}
\noindent
The distributed source coding concept followed throughout this work is characterized by the fact that it can be represented by a simple index assignment stage, i.e. by a one-to-one mapping from the quantization indices $i_n \in \mathcal{I}_n$ to the codewords $w_n \in \mathcal{W}_n$ such that $w_n = m_n(i_n)$ for all $n \in \mathcal{N}$. Considering this low-complexity approach, distributed compression can be achieved by choosing $|\mathcal{W}_n| < |\mathcal{I}_n|$, i.e. whenever we have fewer codewords than quantization levels.\footnote{Such index assignments generally increase the distortion of the system, because information is lost during the mapping process, i.e. more than one quantization index $i_n$ might lead to one and the same codeword $w_n$. However, since the rate can be reduced considerably, this method offers a way to achieve a wider range of rate/distortion trade-offs.}
Thus, the data rate can be reduced from $R_n' = \lceil \log_2|\mathcal{I}_n| \rceil$ to $R_n = \lceil \log_2|\mathcal{W}_n| \rceil$ [bits/sample] for several encoders $n \in \mathcal{N}$. The goal is to jointly design such index assignments such that the  end-to-end distortion $d(\Psi)$ for an arbitrarily chosen subset of sources $\Psi \subseteq \mathcal{N}$ is minimized.
The design procedure presented in the following was inspired by~\cite{FlynnG:87} where a coding solution for two correlated observations was presented. In this work, we generalize the corresponding design algorithm to construct distortion optimized index assignments for an arbitrary subset of encoders $\Omega \subseteq \mathcal{N}$. 
It is worth mentioning that, in principle, several other methods can be used to construct suitable index assignments,
e.g. those based on syndromes~\cite{sandeep-kannan:discus}, on Diophantine index assignments~\cite{maierbacherB:dio} or even on random index assignments.
However, the presented method has the advantage that it is very versatile and that the distortion itself serves as an optimization criterion within the design.

\subsection{Optimization Criterion}
\label{s:distOpt}
\noindent
The figure of merit for our optimization procedure is the minimization of the end-to-end distortion $d(\Psi)$ for an arbitrarily chosen subset of sources $\Psi \subseteq \mathcal{N}$ which, for the case of a MSE distortion metric, can be expressed as follows:
\begin{eqnarray}
d(\Psi)    & = & E\{||\hat{\mathbf{U}}_{\Psi} - \mathbf{U}_{\Psi}||^2\} = E\{(\hat{\mathbf{U}}_{\Psi} - \mathbf{U}_{\Psi})^T \cdot (\hat{\mathbf{U}}_{\Psi} - \mathbf{U}_{\Psi})\} \nonumber \\
	& = & E\{ \sum_{\forall n \in \Psi} (\hat{U}_n - U_n)^2\} = \sum_{\forall n \in \Psi} E\{(\hat{U}_n - U_n)^2\}
\label{eqn:optCriterionIR}
\end{eqnarray}
where $\mathbf{U}_{\Psi}$ denotes the vector of considered source variables $U_n$ and, equally, $\hat{\mathbf{U}}_{\Psi}$ denotes the vector of estimates $\hat{U}_n$ considered within the calculation, $n \in \Psi$. 
In Appendix~\ref{a:dist} it is shown that the distortion associated with each source $d(n) = E\{(\hat{U}_n - U_n)^2\}$, $n \in \Psi$, can be expressed as follows
\begin{equation}
d(n) = E\{(\tilde{U}_n - U_n)^2\} + E\{(\hat{U}_n - \tilde{U}_n)^2\}
\label{eqn:distSum}
\end{equation}
where the reconstruction levels of the quantizers are assumed to be the centroid of the quantization cells such that $\tilde{u}_{n,i_n} = E\{U_n|i_n\}$ for all $i_n \in \mathcal{I}_n$.
We see that the distortion $d(n)$ consists of two components where $d_q(n) = E\{(\tilde{U}_n - U_n)^2\}$
is the component directly resulting from the finite granularity of the scalar quantizer $q_n$, $n \in \Psi$, and 
\begin{equation}
d_d(n) = E\{(\hat{U}_n - \tilde{U}_n)^2\} = \sum_{\forall \mathbf{i}_{\Psi} \in \mathcal{I}_{\Psi}} p(\mathbf{i}_{\Psi}) \cdot (\hat{u}_n(\mathbf{w}_{\mathcal{T}}) - \tilde{u}_{n,i_n})^2
\label{eqn:distD}
\end{equation}
is the component mainly affected by the choice of the estimate $\hat{u}_n(\mathbf{w}_{\mathcal{T}})$ for the vector of available codewords $\mathbf{w}_{\mathcal{T}} \in \mathcal{W}_{\mathcal{T}}$, $\mathcal{W}_{\mathcal{T}}=\prod_{\forall l \in \mathcal{T}}\mathcal{W}_l$, $\mathcal{T}\subseteq\mathcal{N}$. The latter depends directly on the configuration of the index assignments $m_l$ of all encoders $l \in \mathcal{T}$;
compare e.g.~(\ref{eq:estimatesOpt}) for the case where CME is considered for decoding.
For the design of the index assignments, as presented in the following, we set $\mathcal{T}$ equal to $\Omega$, i.e.
the codewords from all encoders $\Omega$ are assumed to be known. 
Based on~(\ref{eqn:distSum}), we can state that the distortion of all sources $n \in \Psi$ can be expressed by the sum
\begin{equation}
d(\Psi) = \sum_{\forall n \in \Psi} d(n) = \sum_{\forall n \in \Psi} (d_q(n)+d_d(n)) = \sum_{\forall n \in \Psi} d_q(n) + \sum_{\forall n \in \Psi} d_d(n)
\label{eqn:distAll}
\end{equation}
where $d_q(\Psi) = \sum_{\forall n \in \Psi} d_q(n)$ is caused by the quantization stage and $d_d(\Psi) = \sum_{\forall n \in \Psi} d_d(n)$ is caused by the index assignment stage.
It is worth pointing out that the calculation of $d_q(\Psi)$ does not take into account any knowledge about the actual configuration of the index assignments which is very helpful for design purposes, as presented next.

\subsection{Index-Reuse Algorithm}
\label{s:indexReuse}
\noindent
The basic idea underlying the presented algorithm is to construct index assignments in an iterative fashion.
In each step of this procedure, the number of output codewords is reduced such that, in general, more than one quantization index is assigned to each codeword index. This means that the codeword indices are reused, while considering the resulting end-to-end distortion as the optimization criterion.

Starting with bijective mappings between the quantization indices $i_n$ and the codewords $w_n$, where the number of codewords is equal to the number of quantization indices, i.e. $|\mathcal{I}_n|=|\mathcal{W}_n|$, the algorithm subsequently modifies the mapping functions $m_n$ for all considered encoders $n \in \Omega$ by merging two codewords (or, equivalently, the originating quantization indices) to a single new codeword. This is repeated until the targeted number of codewords, denoted as $K$, is reached. In each step of the procedure, the algorithm chooses the merging from all possible candidates yielding the minimum distortion $d(\Psi)=d_q(\Psi)+d_d(\Psi)$ calculated for the set of considered sources $\Psi$ 
where only $d_d(\Psi)$ is affected by the index assignments.\footnote{We note that 
the search algorithm is not optimal due to the single-step nature of the optimization.}

For a detailed discussion of the algorithm, we assume that $|\mathcal{I}_n| = L$ and $|\mathcal{W}_n| = L$ in the beginning of the procedure and that $|\mathcal{W}_n| = K$, $K < L$, at the end of the procedure, for all $n \in \Omega$.
For implementation purposes, we assume that $\mathcal{W}_n = \{0,1,\sldots,|\mathcal{W}_n|-1\}$ and represent the mapping functions $m_n:~~~\mathcal{I}_n \rightarrow \mathcal{W}_n$ by vectors $\mathbf{f}_n \in \mathcal{W}_n^{|\mathcal{I}_n|}$ such that the codeword $w_n$ can be obtained from the index $i_n$ by simple vector referencing where $w_n = \mathbf{f}_n(i_n)$ for all $i_n \in \mathcal{I}_n$ and for all $n \in \Omega$.
The merging of two codewords $w_n = a$ and $b$ within the vector $\mathbf{f}_n$ shall be described by the merging function $g:~~~ \mathcal{W}_n^{|\mathcal{I}_n|} \times \mathcal{W}_n \times \mathcal{W}_n \rightarrow \mathcal{V}_n^{|\mathcal{I}_n|}$ where $\mathcal{V}_n \in \{0,1,\sldots,|\mathcal{W}_n|-2\}$ is the resulting codeword alphabet with a reduced number of codewords and the vector $\mathbf{e}_n$, describing the resulting mapping, can be obtained from $\mathbf{f}_n$ by $\mathbf{e}_n = g(\mathbf{f}_n,a,b)$. Assuming that at the initialization of the algorithm the vector $\mathbf{f}_n$ was initialized such that $\mathbf{f}_n = (0,1,\sldots,|\mathcal{I}_n|-1)^T$ and that $a < b$, then it is easy to show that $\mathbf{e}_n$ can be obtained from $\mathbf{f}_n$ by performing the following assignment for $i_n = 0,1,\sldots,|\mathcal{I}_n|-1$:
\begin{equation}
\mathbf{e}_n(i_n) = 
\begin{cases}
    a 				&,\textrm{for}~\mathbf{f}_n(i_n)=a~\textrm{or}~\mathbf{f}_n(i_n)=b\\
    \mathbf{f}_n(i_n)-1 	&, \textrm{for}~\mathbf{f}_n(i_n) > b\\
    \mathbf{f}_n(i_n)		&, \textrm{otherwise}.
\end{cases}
\nonumber
\end{equation}
Let $\mathcal{E}_n = \{\mathbf{f}_m: m \in \Omega, m \neq n\} \cup \{\mathbf{e}_n\}$ be the collection of mapping vectors after merging two codewords in $\mathbf{f}_n$. We use the notational convention that $d_d(\Psi,\mathcal{E}_n)$ can be used to indicate that the distortion $d_d(\Psi)$ according to (\ref{eqn:distD}) was calculated based on those mapping functions. A detailed formulation of the whole procedure can be found in Algorithm~\ref{alg:indexreuse}. 
\begin{algorithm}
\dontprintsemicolon
{\bf Initialization}\\
$\bullet$ start with one-to-one mapping\\
\qquad $\mathbf{f}_n \leftarrow (0,1,\sldots,L-1)^T$, for all $n \in \Omega$\;
$\bullet$ set initial number of codewords\\
\qquad $k \leftarrow L$\;
{\bf Main Loop}\\
\While{($k > K$)}{
	\For{($n \in \Omega$)}{
		$\bullet$ set reference distortion to maximum\\
		\qquad $d^{*} \leftarrow \infty$\;
		\For{($a=0,1,\sldots,k-2$)}{ 
			\For{($b=a+1,a+2,\sldots,k-1$)}{
				$\bullet$ merge cell $a$ and cell $b$ within $\mathbf{f}_n$\\
				\qquad $\mathbf{e}_n=g(\mathbf{f}_n,a,b)$\;
				$\bullet$ calculate resulting overall distortion\\
				\qquad $d=d_d(\Psi,\mathcal{E}_n)$\;
				\If{($d<d^{*}$)}{
					$\bullet$ save current mapping and distortion\\
					\qquad $d^{*} \leftarrow d$\;
					\qquad $\mathbf{f}_n \leftarrow \mathbf{e}_n$\;
				}
			}
		}
	}
	$\bullet$ reduce number of codewords by one\\
	\qquad $k \leftarrow k-1$\\
	}
\caption{Index-Reuse Optimization Algorithm\label{alg:indexreuse}}
\end{algorithm}

Since this particular property was required in the problem setup of Section~\ref{s:setup}, it is worth pointing out that the presented algorithm constructs index assignments that are surjective functions. In the initial step of the algorithm the index assignments are assumed to be bijective functions which, by definition, are also surjective. In each further step of the procedure two codewords in the original mapping are mapped (merged) onto a single new codeword. It is easy to see that this corresponds to the case where the indices that were mapped to either one of the original codewords are now mapped onto the newly created codeword. Thus, the assignment is still a function, since the involved indices are still mapped onto a codeword, and it is also surjective, since for the newly created codeword there always exist some indices that are mapped onto it. This is valid for each step of the procedure and, by induction, the mapping created after any number of steps is (still) a surjective function.

For the important case where the set of considered sources is equal to the set of considered encoders, i.e. when $\Psi=\Omega$, the complexity of the optimization algorithm is discussed in detail in Appendix~\ref{a:complexityIR}. It is shown that the algorithm can be implemented with a computational complexity that grows exponential with $|\Omega|$ making it feasible only for a small number of encoders $|\Omega|$. A reasonable way to decrease the overall complexity for a large number of encoders is to form clusters of encoders and optimize each cluster separately, as explained in the next section.

\section{Source-Optimized Clustering}
\label{s:sourceModel}
\noindent
The need for a computationally feasible code design motivates us to partition the entire set of encoders 
into subsets (clusters). The encoders can then be optimized within each cluster, thus, reducing the optimization effort for the encoding side. Moreover, this clustering and coding strategy can be easily combined with the scalable decoder presented in~\cite{bar05} which relies on a carefully chosen factor graph model and allows for joint decoding of the data sent by {\it all} encoders. The key towards computationally feasible joint decoding is for the decoder to use an approximated PDF $\hat{p}(\mathbf{u})$ instead of
$p(\mathbf{u})$ as basis for efficient decoding considering only the statistical dependencies within certain subsets of sources. 
Therefore, it becomes crucial to build the decoding model and the source clusters alongside to ensure that statistical dependencies, which are exploited during encoding to reduce redundancy within the clusters, are still available at the decoder to compensate for the information loss imposed by the index assignment stage.  
In \cite{bar05} the Kullback-Leibler distance (KLD) was deemed to be a suitable measure to estimate the impact of the chosen  decoding model onto the overall system performance, i.e. the MSE distortion. 
Since we are interested in minimizing the overall system MSE, we chose the KLD as optimization criterion to find not only a suitable source approximation but also adequate clusters.

\subsection{Preliminaries}\label{s:pre}
\noindent
The PDF $p(\mathbf{u})$ can be approximated by assuming a factorization of the form $\hat{p}(\mathbf{u}) =
\prod_{m = 1}^M f_m(\mathbf{u}_{\mathcal{S}_m})$
where $\mathcal{S}_m \subseteq \mathcal{N}$ for $m = 1,2,\sldots,M$
are subsets of source indices such that
$\bigcup_{m = 1}^M \mathcal{S}_m = \mathcal{N}$. Since generally
$p(\mathbf{u}) \neq \hat{p}(\mathbf{u})$, the resulting PDF
$\hat{p}(\mathbf{u})$ is an approximation of
$p(\mathbf{u})$.

Specifically, we shall consider constrained chain rule expansions (CCREs) of $p(\mathbf{u})$ that can be obtained
from the regular chain rule expansion by removing some of the conditioning
variables.
More formally, a factorization
\begin{equation}
\hat{p}(\mathbf{u}) = \prod_{m = 1}^M f_m(\mathbf{u}_{\mathcal{S}_m}) =
\prod_{m = 1}^{M}p(\mathbf{u}_{\mathcal{A}_m}|\mathbf{u}_{\mathcal{B}_m}),
\label{eqn:CCREfactorization}
\end{equation}
where $\mathcal{A}_m$, $\mathcal{B}_m$ and $\mathcal{S}_m = \mathcal{A}_m \cup \mathcal{B}_m$
are subsets of the elements in $\mathcal{N}$, is a
CCRE of $p(\mathbf{u})$, if the following constraints are met for $m = 1,2,\sldots,M$:
\begin{equation}
\mathcal{A}_m \cap \mathcal{B}_m = \emptyset, 
~~~~ \bigcup_{m = 1}^M \mathcal{A}_m = \mathcal{N}, 
~~~~ \mathcal{B}_m \subseteq \bigcup_{l=1}^{m-1} \mathcal{A}_l.
\label{eqn:CCREConditions}
\end{equation}
Notice that the set $\mathcal{B}_1$ is always empty and that $\mathcal{B}_m=\bigcup_{l=1}^{m-1}\mathcal{A}_l$ holds for the usual chain rule expansion.
We call a CCRE {\it symmetric} if any $\mathcal{B}_m$ with $m = 2,3,\cdots,M$
is a subset of $\mathcal{S}_l$ for some $l<m$.

The Kullback-Leibler distance (KLD) between a PDF $p(\mathbf{u})$
and its approximation $\hat{p}(\mathbf{u})$ is defined as
\begin{equation}
D(p(\bf u)||\hat{p}(\bf u))
=\idotsint p(\bf u) \log_2 \frac{p(\bf u)}{\hat{p}(\bf u)} \;d \bf u,
\label{eqn:KLD1}
\end{equation}
e.g. see~\cite{cover-thomas:it-book}, which can be used as
optimization criterion when constructing source factorizations.
In \cite{bar05} it was shown that the KLD can be calculated
explicitly for CCREs of Gaussian
PDFs $\mathcal{N}(\mathbf{0}_N,\mat{R})$ as follows
\begin{equation}
D(p(\bf u)||\hat{p}(\bf u))
= -\frac{1}{2}\log_2|\mat{R}|+ \sum_{m = 1}^M 
\Delta D(\mathcal{S}_m,\mathcal{B}_m)
\label{eqn:KLD2}
\end{equation}
where the KLD benefit obtained by introducing the factor 
$p(\mathbf{u}_{\mathcal{A}_m} | \mathbf{u}_{\mathcal{B}_m})$ is given by
\begin{equation}
\Delta D(\mathcal{S}_m,\mathcal{B}_m)= \frac{1}{2} \log_2 \frac{|\mat{R}_{\mathcal{S}_m}|}{|\mat{R}_{\mathcal{B}_m|}}
\label{eqn:deltaKLD}
\end{equation}
where $\mat{R}_{\mathcal{S}_m}$ as well as $\mat{R}_{\mathcal{B}_m}$ are
the covariance matrices of the Gaussian PDFs
$p(\mathbf{u}_{\mathcal{S}_m})$ and
$p(\mathbf{u}_{\mathcal{B}_m})$, respectively.
%

It is worth pointing out that a source factorization according to~(\ref{eqn:CCREfactorization}) can be used directly for an efficient decoder implementation as discussed in Appendix~\ref{a:subOptDec}. In particular, this holds for the case where the number of variables in the factors is bounded such that $|\mathcal{S}_m| \leq S$ for $m = 1,2,\sldots,M$. A complexity analysis for scalable decoding based these assumptions can be found in Appendix~\ref{a:subOptDec}. It is shown that the computational complexity for the case where $|\mathcal{B}_m| = 1$ for $m = 1,2,\sldots,M$ is of $\mathcal{O}(MS\mapCoeff^{S})$. In the other cases the computational complexity is of $\mathcal{O}(TMS\mapCoeff^{S})$ where $T>1$ specifies the maximum number of iterations used for decoding.
Notice that $M$, the number of factors in the factorization, is considered as a parameter here. 
However, it shall be shown later in this work that $M \leq 2N+1$ holds. 

\subsection{Clustering Algorithm}\label{s:clusteringAlgorithm}
\noindent
The clustering algorithm described in the following is based on the principles of hierarchical clustering~\cite{jain99:clustering} and can be seen as a variant of the Ward algorithm \cite{Ward}. 
The goal is to cluster the set of sources $\mathcal{N}$ into subsets $\Lambda_c \subseteq \mathcal{N}$ such that $\bigcup_{\forall c \in \Gamma} \Lambda_c = \mathcal{N}$ and $\Lambda_i \cap \Lambda_j = \emptyset$ for all $i \neq j$ with $\{i,j\} \in \Gamma$ where $\Gamma = \{1,2,\sldots,C\}$ is the set of cluster indices and $C = |\Gamma|$ is the number of clusters. The maximum cluster size $S$ is assumed to be given and defined such that $|\Lambda_c| \leq S$ for all $c \in \Gamma$.

The clusters itself are constructed by a successive merging process. 
The algorithm starts with a set of single-element clusters such that $\Lambda_s' = \{s\}$ for all $s \in \Gamma'$ where $\Gamma' = \mathcal{N} = \{1,2,\sldots,N\}$ is the initial set of cluster indices. 
In each of the following steps two of those clusters are selected and merged into a new cluster. 
The clusters are selected using the KLD $D(p(\mathbf{u})||\tilde{p}(\mathbf{u}))$ between the original PDF $p(\mathbf{u})$ and the approximated PDF $\tilde{p}(\mathbf{u}) = \prod_{\forall s \in \Gamma'} p(\mathbf{u}_{\Lambda_s'})$ as an {\it objective} function where $\tilde{p}(\mathbf{u})$ directly results from the current choice of clusters and $D(p(\mathbf{u})||\tilde{p}(\mathbf{u}))$ is defined analog to~(\ref{eqn:KLD1}). 
For each possible pair of clusters $(\Lambda_k',\Lambda_l')$ with $k \neq l$ and $\{k,l\} \in \Gamma'$, the algorithm determines the current value of the objective function to find the pair $(\Lambda_k',\Lambda_l')$ leading to the smallest KLD between original and approximated PDF. The indices of the selected clusters $(k,l)$ are then removed from the current set of cluster indices $\Gamma'$ whereas the index of the newly created cluster $r$ is added to it.
This procedure is repeated until only a single cluster remains and a history of all mergings performed during the different stages of the optimization procedure is obtained.

Using (\ref{eqn:KLD2}), it is possible to show that the overall KLD can be calculated as follows
\begin{equation}
D(p(\mathbf{u})||\tilde{p}(\mathbf{u})) =
- \frac{1}{2}\log_2 |\mat{R}| + 
\sum_{\forall s \in \Gamma'} \Delta D(\Lambda_s',\emptyset),
\label{eqn:objfun}
\end{equation}
where $\Delta D(\Lambda_s',\emptyset)$ is the KLD benefit imposed by an arbitrary cluster $\Lambda_s'$.
Since the objective function has to be evaluated many
times during the optimization process, it is useful to express~(\ref{eqn:objfun})
in terms of intermediate results to reduce computational complexity.
The differential KLD benefit created by merging an arbitrary pair of clusters
$(\Lambda_k',\Lambda_l')$ with $k \neq l$ and $\{k,l\} \in \Gamma'$ into a new cluster can be expressed as follows
\begin{equation}
\Delta D'(\Lambda_k',\Lambda_l') = \Delta D(\Lambda_k' \cup \Lambda_l',\emptyset) -  \Delta D(\Lambda_k',\emptyset) - \Delta D(\Lambda_l',\emptyset),
\label{eqn:diffKLD}
\end{equation}
which can be used to locally evaluate the impact of the considered merging
onto the overall KLD given by (\ref{eqn:objfun}). 
Assuming that $t$ is the number of mergings performed at a certain stage of the procedure, then the expression 
\begin{equation}
 D(p(\mathbf{u})||\tilde{p}(\mathbf{u})) =
- \frac{1}{2}\log_2 |\mat{R}| + \sum_{s=1}^t \Delta D'(\Lambda_{k(s)}',\Lambda_{l(s)}')
\nonumber
\end{equation}
can be used to evaluate the overall KLD in (\ref{eqn:objfun}) based on the differential KLD benefits in~(\ref{eqn:diffKLD}) only.

\begin{figure}[tb]
\centerline{
\begin{tabular}{cc}
\begin{minipage}{6.3cm}
\includegraphics[width=6.5cm]{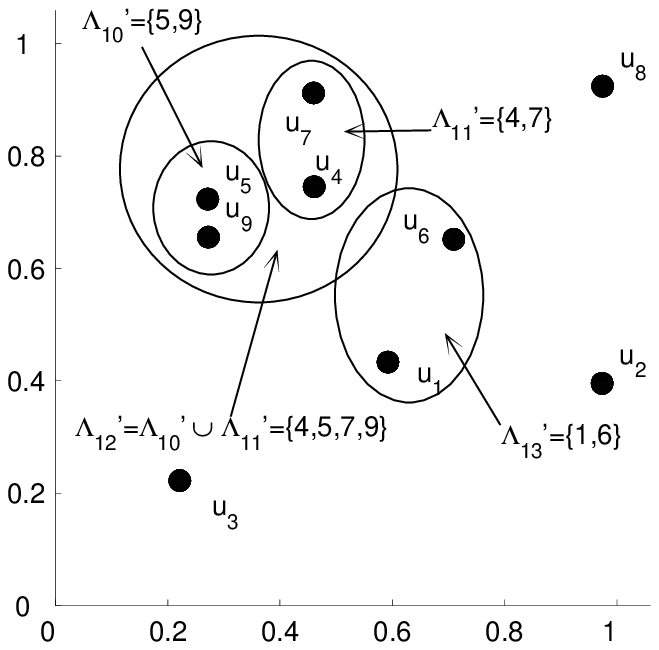}
\end{minipage}
&
\begin{minipage}{5.7cm}
\includegraphics[width=5.7cm]{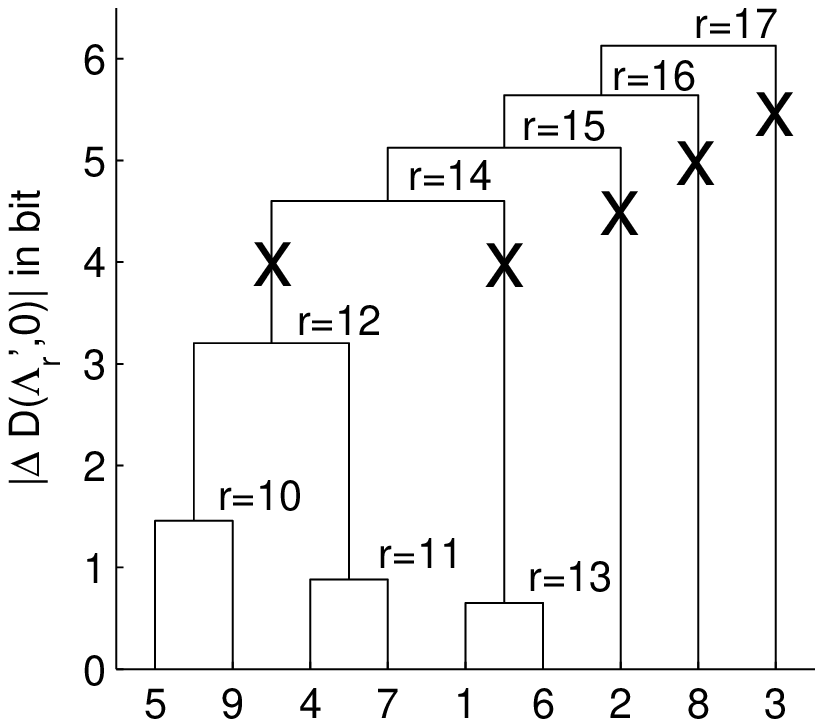}
\end{minipage}
\\
(a)
&
(b)
\\
\end{tabular}
}
\caption{Example. Source-optimized clustering procedure with $N=9$
uniformly distributed sensors picking-up observations $u_1,...,u_9$:
(a) Mergings performed during the hierarchical clustering procedure for the first
four iteration steps leading to resulting clusters $\Lambda_r'$
with indices $r=10,\cdots,13$.
(b) Tree representation of the mergings performed during different the
stages of the optimization process (dendrogram).
The KLD benefit $|\Delta D(\Lambda_r',\emptyset)|$ in [bit] imposed by the clusters $\Lambda_r'$
is provided for all iteration steps.
The branches of the tree that are cut during the pruning process with a maximum cluster size of $S=4$ are marked with a cross.
After pruning, the source clusters $\Lambda_1=\{3\}$, $\Lambda_2=\{8\}$, $\Lambda_3=\{2\}$, $\Lambda_4=\{1,6\}$ and $\Lambda_5=\{4,5,7,9\}$ can be defined.
}
\label{fig:clusteringExample}
\end{figure}

A detailed description of the entire procedure can be found in Algorithm \ref{alg:clustering} where 
$r$ labels the clusters in ascending order and $h$ (a two-dimensional array) is used to store a history of the mergings performed during different stages of the clustering procedure. 
In Figure \ref{fig:clusteringExample}(a) the merging process is illustrated for an exemplary scenario. A graphical representation of the mergings performed during different stages of the optimization, the so-called dendrogram \cite{jain99:clustering}, is shown in Figure \ref{fig:clusteringExample}(b).
\begin{algorithm}
\dontprintsemicolon
{\bf Initialization}\\
$\bullet$ start with one-element clusters\\
\qquad $\Gamma' \leftarrow \{1,...,N\}$\;
\qquad $\Lambda_s' \leftarrow \{s\}$, for all $s \in \Gamma'$\;
\qquad $t \leftarrow 1,~r \leftarrow N+t$\;
{\bf Main Loop}\\
\Repeat{($|\Gamma'|=1$)}{
    	$\bullet$ find the pair of cluster $(\Lambda_k',\Lambda_l')$ with $k \neq l$ and $\{k,l\} \in \Gamma'$\\
	such that $|\Delta D'(\Lambda_k',\Lambda_l')|$ is maximized\;
    	$\bullet$ store intermediate results: \\
	\qquad $\Lambda_r' \leftarrow \Lambda_k' \cup \Lambda_l'$\;
    	$\bullet$ delete original clusters from index list: \\
	\qquad $\Gamma' \leftarrow \Gamma' \backslash \{k,l\}$\;
    	$\bullet$ add new cluster to index list: \\
	\qquad $\Gamma' \leftarrow \Gamma' \cup \{r\}$\;
    	$\bullet$ save clustering history: \\
	\qquad $h(t,1) \leftarrow k,~h(t,2) \leftarrow l$\;
    	$\bullet$ update internal variables: \\
	\qquad $t \leftarrow t+1,~r \leftarrow N+t$\;
	}
\caption{KLD optimized clustering\label{alg:clustering}}
\end{algorithm}
%

Using the dendrogram derived before, the source clusters $\Lambda_c$ with a maximum cluster size of $S$ can be constructed.
We start at the root of the dendrogram, which is basically a tree, and 
descend along its branches to lower hierarchical levels. 
While moving from one level to the next lower one, the dendrogram branches 
into two subtrees. The number of leafs, i.e. the number of sources
connected to each subtree are counted and if the number of leafs of one (or both)
subtree(s) is smaller or equal to $S$, we cut the corresponding
subtree out of the dendrogram. This pruning process is
repeated until all leafs are removed from the dendrogram. When
the pruning is finished, the subtrees are labeled by
the successively increased index $c=1,2,\sldots,C$.
The source clusters $\Lambda_c$, $c \in \Gamma$, can then be determined from the subtrees by
assigning the variables $n \in \mathcal{N}$ (associated with each of the subtree's leafs)
to the corresponding cluster.
The overall KLD $D(p(\mathbf{u})||\check{p}(\mathbf{u}))$ between the original PDF $p(\mathbf{u})$ and the approximated PDF $\check{p}(\mathbf{u}) = \prod_{\forall c \in \Gamma} p(\mathbf{u}_{\Lambda_c})$ can then be calculated based on the resulting clusters 
\begin{equation}
D(p(\mathbf{u})||\check{p}(\mathbf{u})) =
- \frac{1}{2}\log_2 |\mat{R}| + 
\sum_{\forall c \in \Gamma} \Delta D(\Lambda_c,\emptyset)
\label{eqn:KLDClusterFinal}
\end{equation}
where $D(p(\mathbf{u})||\check{p}(\mathbf{u}))$ is defined analog to~(\ref{eqn:KLD1}).
In Figure \ref{fig:clusteringExample}(b) the pruning process is illustrated for the previous example. 

Because of the hierarchical merging concept based on local decisions, 
the proposed clustering algorithm is in general sub-optimal.
However, the hierarchical approach has the advantage that the 
resulting dendrogram can be used elegantly to construct clusters
with a bounded number of source variables $S$.\footnote{With partitional clustering methods, see e.g.~\cite{jain99:clustering}, this would be an arguably difficult task.}

In Appendix~\ref{a:cluster} it is shown that the computational complexity of source-optimized clustering (evaluated in a very pessimistic fashion) is of $\mathcal{O}(N^5 \log N)$ 
which makes the overall procedure feasible for medium to large values of $N$.
Furthermore, it is easy to show that the number of clusters $C$ with a maximum cluster size of $S$ is bounded according to $C \leq \lfloor \frac{N}{S}\rfloor$ which shall be required in the next section.

\subsection{Source-Optimized Factorization}\label{s:clusterTrees}
\noindent
In the last section we have shown how to construct KLD optimized clusters fitting our purposes.
The second step towards our goal of obtaining a source factorization
is to transduce the derived clusters into a symmetric CCRE of the form
$\hat{p}(\mathbf{u}) = \prod_{m = 1}^M f_m(\mathbf{u}_{\mathcal{S}_m})$ matching the conditions in~(\ref{eqn:CCREConditions}).
This can be achieved by {\it linking} the clusters $\Lambda_c$,
$c \in \Gamma$, successively together.

The basic principle of the linking procedure is as follows: After choosing a specific cluster 
as starting point for the procedure, select one of the unconnected clusters
(i.e. a cluster which is not yet considered in the source factorization) and link
it with the already connected clusters (i.e. incorporate it into the source factorization).
Assuming that cluster $r \in \Gamma$ was chosen as the starting point for the 
optimization, we can define a set of linked clusters $\Gamma' = \{r\}$ 
and a set of unconnected clusters $\overline{\Gamma'} = \Gamma \backslash \{r\}$. 
At each step of the procedure a cluster $k \in \Gamma'$ and a cluster $l \in \overline{\Gamma'}$ are selected.
The index $l$ is added to the set of linked clusters, i.e. $\Gamma' = 
\Gamma' \cup \{l\}$, and removed from the set unconnected clusters, i.e. $\overline{\Gamma'} = 
\overline{\Gamma'} \backslash \{l\}$. This is repeated until all clusters are linked, 
i.e. $|\Gamma'| = |\Gamma|$. 

More specifically, two clusters $\{k,l\} \in \Gamma$ are linked by choosing a set of variables $\mathcal{P}_k
\subseteq \Lambda_k$ and a set of variables $\mathcal{Q}_l \subseteq \Lambda_l$. These sets
will form the basis of the factor introduced into the source factorization.
Since the complexity of scalable decoding is highly dependent on the number
of variables within the single factors of the underlying source factorization (see Appendix~\ref{a:subOptDec} or~\cite{bar05} for details),
we introduce the design parameters $A$ and $B$
such that $|\mathcal{P}_k| \leq A$ and $|\mathcal{Q}_l| \leq B$
for all $\{k,l\} \in \Gamma$.

The source factorization starts with a single 
factor $p(\mathbf{u}_{\mathcal{I}_r})$ containing the variables of the initially chosen cluster $r$, i.e. $\mathcal{I}_r = \Lambda_r$.
While establishing a link between the two clusters $k$ and $l$,
the factors $p(\mathbf{u}_{\mathcal{Q}_l}|\mathbf{u}_{\mathcal{P}_k})$ and
$p(\mathbf{u}_{\mathcal{I}_l}|\mathbf{u}_{\mathcal{Q}_l})$ are added to the 
source factorization where $\mathcal{I}_l = \Lambda_l \backslash \mathcal{Q}_l$.
As the clusters are linked, a source factorization is constructed step-by-step
where the running index $d$ is used to index the added clusters. 
The resulting source factorization can then be written as 
\begin{equation}
\hat{p}(\mathbf{u}) = \underbrace{p(\mathbf{u}_{\Lambda_{l(d=1)}})}_{\textrm{(a)}}
\prod_{d=2}^{C} \Big( \underbrace{p(\mathbf{u}_{\mathcal{Q}_{l(d)}} | \mathbf{u}_{\mathcal{P}_{k(d-1)}})}_{\textrm{(b)}}
\underbrace{p(\mathbf{u}_{\mathcal{I}_{l(d)}} | \mathbf{u}_{\mathcal{Q}_{l(d)}})}_{\textrm{(c)}} \Big)
\label{eqn:factorization}
\end{equation}
where $\Lambda_{l(d=1)} = \Lambda_r = \mathcal{I}_{l(d=1)} = \mathcal{I}_r$ and all factors with $\mathcal{I}_{l(d)} = \emptyset$ for $d = 2,\cdots,C$ are discarded. 
Notice that, when constructed according to the aforementioned linking procedure, 
there exists a one-to-one correspondence between the running index $d$ and the cluster 
indices $c \in \Gamma$. Moreover, it can easily be shown that~(\ref{eqn:factorization}) fullfills the 
criteria of CCREs by verifying the conditions in~(\ref{eqn:CCREConditions}).

After discussing how source factorizations fitting our purposes can be constructed,
we are ready to show how to choose the subsets $\mathcal{P}_k$ and $\mathcal{Q}_l$ and in which order the clusters should be linked such that the overall KLD $D(p(\mathbf{u})||\hat{p}(\mathbf{u}))$ defined analog to~(\ref{eqn:KLD1}) is minimized.

It is easy to show that the KLD of the source factorization given in~(\ref{eqn:factorization}) can be expressed as
\begin{eqnarray}
D(p(\mathbf{u})||\hat{p}(\mathbf{u})) & = & -\frac{1}{2}\log_2|\mat{R}| + \Delta D(\Lambda_{l(d=1)},\emptyset)  \nonumber \\
& & +\sum_{d=2}^C \Big( \Delta D(\mathcal{P}_{k(d-1)} \cup \mathcal{Q}_{l(d)},\mathcal{P}_{k(d-1)})
+ \Delta D(\Lambda_{l(d)},\mathcal{Q}_{l(d)}) \Big).
\label{eqn:KLDFactorization}
\end{eqnarray}
The KLD benefit imposed by the factors~(c) in~(\ref{eqn:factorization}) can be written
\begin{eqnarray}
\Delta D(\Lambda_l,\mathcal{Q}_l) 
    &= & \frac{1}{2} \log_2 \frac{|\mat{R}_{\Lambda_l}|}{|\mat{R}_{\mathcal{Q}_l}|} = \frac{1}{2} \log_2 |\mat{R}_{\Lambda_l}| - \frac{1}{2} \log_2 |\mat{R}_{\mathcal{Q}_l}| \nonumber \\
  & = & \Delta D(\Lambda_l,\emptyset) - \Delta D(\mathcal{Q}_l,\emptyset)
\label{eqn:KLDFactor1}
\end{eqnarray}
since the covariance matrices $\mat{R}_{\Lambda_l}$ and $\mat{R}_{\mathcal{Q}_l}$ are symmetric and positive-semidefinite and, thus, the determinants $|\mat{R}_{\Lambda_l}|$ and $|\mat{R}_{\mathcal{Q}_l}|$ are 
non-negative.
Similarly, the KLD benefit imposed by the factors~(b) in~(\ref{eqn:factorization}) can be expressed as
\begin{equation}
\Delta D(\mathcal{P}_k \cup \mathcal{Q}_l, \mathcal{P}_k) = \Delta D(\mathcal{P}_k \cup \mathcal{Q}_l, \emptyset) -
\Delta D(\mathcal{P}_k, \emptyset).
\label{eqn:KLDFactor2}
\end{equation}
Considering the KLD benefit in~(\ref{eqn:KLDFactor1}) and~(\ref{eqn:KLDFactor2}), we notice that $\Delta D(\Lambda_l, \emptyset)$ in~(\ref{eqn:KLDFactor1}) already was considered during the cluster optimization in Section~\ref{s:sourceModel} . Thus, we are able to define the KLD benefit of establishing a link based on the sets $\mathcal{P}_k$ and $\mathcal{Q}_l$ as 
\begin{equation}
\Delta D^{\ast}(\mathcal{P}_k,\mathcal{Q}_l) = \Delta D(\mathcal{P}_k \cup \mathcal{Q}_l, \emptyset) 
- \Delta D(\mathcal{P}_k,\emptyset) - \Delta D(\mathcal{Q}_l,\emptyset).
\label{eqn:KLDLink}
\end{equation}
Using~(\ref{eqn:KLDClusterFinal}), the KLD of the source factorization in~(\ref{eqn:factorization}) can be written as
\begin{equation}
D(p(\mathbf{u})||\hat{p}(\mathbf{u})) = D(p(\mathbf{u})||\check{p}(\mathbf{u})) 
  + \sum_{d=2}^C \Delta D^{\ast}(\mathcal{P}_{k(d-1)},\mathcal{Q}_{l(d)}),
\label{eqn:decouplingKLD}
\end{equation}
decoupling the link optimization from the cluster optimization.

If an already linked cluster $k$ is to be connected to a cluster $l$ in a KLD optimal way, then the sets $\mathcal{P}_k \subseteq \Lambda_k$ and $\mathcal{Q}_l \subseteq \Lambda_l$ have to be chosen such that the KLD benefit $\Delta D^{\ast}(\mathcal{P}_k,\mathcal{Q}_l)$ according to (\ref{eqn:KLDLink}) is maximized in magnitude.
The set of all possible subsets $\mathcal{P}_k' \subseteq \Lambda_k$ with $|\mathcal{P}_k'| = A$ is denoted
as $\mathcal{T}(A,\Lambda_k)$ and the set of all possible subsets $\mathcal{Q}_l' \subseteq \Lambda_l$ with $|\mathcal{Q}_l'| = B$ is denoted as $\mathcal{T}(B,\Lambda_l)$. $\mathcal{P}_k$ and $\mathcal{Q}_l$ are therefore defined as
\begin{equation}
(\mathcal{P}_k,\mathcal{Q}_l) = \argmin_{
\begin{smallmatrix}
 (\mathcal{P}_k',\mathcal{Q}_l'): & \mathcal{P}_k' \in \mathcal{T}(A,\Lambda_k) \\
                                  & \mathcal{Q}_l' \in \mathcal{T}(B,\Lambda_l)
\end{smallmatrix}
} 
\Big\{ \Delta D^{\ast}(\mathcal{P}_k',\mathcal{Q}_l') \Big\}
\label{eqn:linkSet}
\end{equation}
and we define link cost as
\begin{equation}
c_{k,l} =  \Delta D^{\ast}(\mathcal{P}_k,\mathcal{Q}_l).
\label{eqn:cost}
\end{equation}
Notice that generally $c_{k,l} \neq c_{l,k}$.

To determine how the clusters are to be linked (i.e. which clusters are to be linked and in which direction), a graph can be constructed representing the KLD optimal links between the clusters. 
The vertices of the graph are obtained by contracting each cluster $\Lambda_c$, with $c \in \Gamma$, to a single vertex $v_c$ and defining the set of vertices as
\begin{equation}
\mathcal{V} = \{v_c: c \in \Gamma\}.
\nonumber
\end{equation}
The set of all possible directed edges $e_{k,l} = (v_k,v_l)$ between the vertices $v_k$ and $v_l$, $\{k,l\} \in \Gamma$, is defined as 
\begin{equation}
\mathcal{E} = \{e_{k,l}=(v_k,v_l): \{k,l\} \in \Gamma, k \neq l \},
\nonumber
\end{equation}
where the cost $c_{k,l}$ of each edge $e_{k,l}$ in terms of KLD benefit is given by (\ref{eqn:cost}). A fully connected graph $\mathcal{G} = (\mathcal{V},\mathcal{E})$ is thus obtained. Provided that the clusters are considered fixed, the overall KLD of the source factorization (\ref{eqn:factorization}) can be optimized solely by optimizing the cluster links, please refer to~(\ref{eqn:decouplingKLD}), which are in turn represented by the directed edges in $\mathcal{G}$. The optimization problem therefore reduces to the Minimum (cost) Directed Spanning Tree (MDST) problem
for which first algorithms were found by Chu and Liu~\citeyear{ChuLiu:MDST} as well as by Edmonds~\citeyear{Edmonds:MDST} to be generalized later by Georgiadis~\citeyear{Georgiadis:MDST}. After applying one of these algorithms to the fully connected graph $\mathcal{G}$, the MDST $\mathcal{G}' = (\mathcal{V},\mathcal{E}')$ with $\mathcal{E}'\subseteq \mathcal{E}$ and its root vertex (i.e. the vertex $v_r \in \mathcal{V}$ which only has outgoing edges) can be found. 
The source factorization (\ref{eqn:factorization}) can then be constructed by moving along the edges of the obtained tree $\mathcal{G}'$ (possibly inspired by a Depth-First Search, see e.g.~\cite[p. 484]{AhoUllman:Foundations}) and linking the clusters corresponding to the visited vertices together. More specifically, the root vertex of $\mathcal{G}'$ corresponds to the factor denoted as~(a) in (\ref{eqn:factorization}), the visited edges correspond to the factors denoted as~(b) and the visited vertices correspond to the factors denoted as~(c). Notice that this tree-based linking approach also conforms with the aforementioned linking procedure, which requires that links result only from already connected clusters, and thus guarantees a valid CCRE.

\begin{figure}[tb]
\centerline{
\begin{minipage}{6.5cm}
\includegraphics[width=6.5cm]{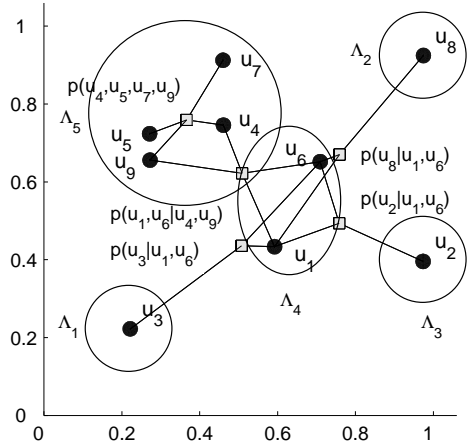}
\end{minipage}
}
\caption{Example. KLD optimized source factorization obtained 
by linking clusters with $A=B=2$. The factor graph represents the
symmetric CCRE $\hat{p}(\mathbf{u}) = p(u_4,u_5,u_7,u_9) \cdot p(u_1,u_6|u_4,u_9)
\cdot p(u_2|u_1,u_6) \cdot p(u_8|u_1,u_6) \cdot p(u_3|u_1,u_6)$.
}
\label{fig:linkingExample}
\end{figure}

Considering the previous example with clusters $\Lambda_1=\{3\}$, $\Lambda_2=\{8\}$, $\Lambda_3=\{2\}$,
$\Lambda_4=\{1,6\}$, $\Lambda_5=\{4,5,7,9\}$ and $A=B=2$, we get the MDST $\mathcal{G}'=(\mathcal{V}, \mathcal{E}')$ with $\mathcal{V} = \{v_1,\cdots,v_9\}$, root $v_5$ and $\mathcal{E}' = \{(v_5,v_4),(v_4,v_2),(v_4,v_3),(v_4,v_1)\}$. Figure \ref{fig:linkingExample} shows the corresponding source factorization.

Appendix~\ref{a:link} discusses the complexity of the source-optimized linking procedure and shows that the computational complexity 
grows exponentially with $S$
assuming that $A = B = \frac{S}{2}$, which makes the overall procedure feasible for small cluster sizes $S$. Notice that in the last section it was shown that $C \leq \lfloor \frac{N}{S}\rfloor$ allowing us to represent the complexity only based on the system parameters $N$ and $S$.
It is easy to see that, because the linking procedure basically constructs a tree between the clusters, the number of factors $M$ in the factorization~(\ref{eqn:CCREfactorization}) can be bounded according to $M \leq  2N+1$. This also means that the number of factors with a maximum size of $S$ is at most $2N+1$, as used in Section~\ref{s:pre} to analyze the complexity of the scalable decoder.
%

\section{Results and Discussion}\label{s:discussion}
\noindent
To underline the effectiveness and efficiency of our low-complexity coding and clustering
strategies, we present numerical performance results for two scenarios with randomly placed sensors and two instances of the so-called CEO Problem~\cite{Berger:quadrGaussianCEO}. 

\subsection{Randomly Placed Sensors}\label{s:generalCase}
\begin{figure}[tb]
\centerline{
\epsfxsize=6.5cm
\epsffile{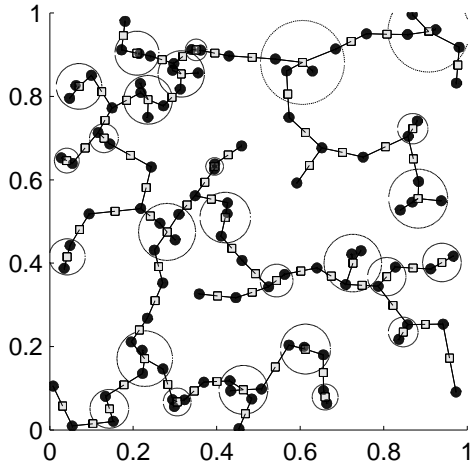}
}
\vspace{-0.2cm}
\caption[Simulation scenario. Graphical representation of the KLD for $N=100$ sensors.]
{Simulation scenario. Graphical representation of the KLD optimized
source factorization for $N=100$ uniformly distributed sensors with correlation
factor $\beta=0.5$. The clusters with a maximum size of $S=4$ (indicated by circles)
were created using the hierarchical clustering method and linked together
by choosing $A=B=1$.}
\label{fig:clustertree}
\end{figure}

\noindent
We consider a unit square with $N=100$ uniformly distributed sensors.
The sensor measurements $u_n$ are Gaussian distributed according to $\mathcal{N}(0,1)$.
As outlined in Section \ref{s:systemm}, we assume that sensor measurements
$\mathbf{u}=(u_1,u_2,\sldots,u_N)^T$ are distributed
according to a multivariate Gaussian distribution $\mathcal{N}({\bf 0}_N,\mat{R})$,
where the correlation between a pair of sensors $u_k$ and $u_l$ decreases
exponentially with the distance $d_{k,l}$ between them, such that
$\rho_{k,l}=\exp(-\beta \cdot d_{k,l})$. Since the performance of our techniques
depend on the correlations between the sensors, we consider
two different scenarios, one with $\beta=0.5$ (strongly correlated sensor
measurements) and one with $\beta=2$ (weakly correlated measurements).
All scalar quantizers at the encoders are Lloyd-Max optimized to minimize the
MSE in the sensor readings $u_n$ using identical resolution
for quantization and identical rates for data transmission,
i.e. $|\mathcal{I}_n| = L$ and $R_n = R$ for all $n\in\mathcal{N}$, $\mathcal{N}=\{1,2,\sldots,N\}$,
where $L \leq 16$ was chosen.
The clusters $\Lambda_c \subseteq \mathcal{N}$ indexed by $c \in \Gamma$ are derived as described in Section~\ref{s:clusteringAlgorithm} where a maximum cluster size of $S=4$ was chosen, see Figure \ref{fig:clustertree}.
The index assignments are then designed successively for all clusters with $|\Lambda_c| > 1$ and $c \in \Gamma$ by employing the IR algorithm described in Section~\ref{s:index} with $\Psi=\Omega=\Lambda_c$. 
Since it is not possible to construct index assignments for single-element clusters, we chose in this case a scalar quantizer (Lloyd-Max optimized as before) with decreased resolution and no index assignments such that $R_n=R$ is still guaranteed for all encoders $n \in \mathcal{N}$.
The source factorization used for decoding is constructed as described in Section~\ref{s:clusterTrees} assuming that $A=B=1$, see Figure~\ref{fig:clustertree}.
The decoder is based on the sum-product algorithm as described in~\cite{bar05} 
where the required PMFs were obtained by Monte Carlo simulation using Lloyd-Max optimized
quantizers with resolution $L_n=L$ for all $n\in\mathcal{N}$.
To evaluate the performance of the coding strategies, we measure the output
signal-to-noise ratio~(SNR) given by
\begin{equation}
\textrm{Output SNR} = 10 \cdot \log_{10}\left(\frac{||\mathbf{u}||^2}{||\mathbf{u}-
\hat{\mathbf{u}}||^2}\right) \textrm{in dB}
\nonumber
\end{equation}
averaged over a $N\times10000$ source samples.
\begin{table}
\caption{Simulation results for $N=100$ and $\beta=\{0.5,~2\}$}
\begin{center}
\begin{tabular}{|l|c|c|c|c|c|c|c|c|}
\hline
					&\multicolumn{4}{|c|}{$\beta = 0.5$}		&
					\multicolumn{4}{|c|}{$\beta = 2$}		\\
$R$ [bit]				&1	&2	&3	&4	&
					1	&2	&3	&4	\\
\hline
$\textrm{SNR}_{\textrm{Dec}}$[dB]	&4.44	&9.46	&14.61	&20.32	&
					4.45	&9.32	&14.65	&20.27	\\
\hline
$\textrm{SNR}_{\textrm{IR}}$[dB]	&11.07	&14.86	&18.29	&N.A.	&
					7.54	&11.72	&16.21	&N.A.	\\
\hline
\end{tabular}
\end{center}
\label{tab:results}
\vspace{-0.4cm}
\end{table}
The simulation results of our system are depicted in Table~\ref{tab:results} for strongly and weakly 
correlated sources. In both scenarios, we consider
the performance achieved when using scalar quantization alone at
the encoder, i.e. where the performance is mainly governed by the
properties of the decoder~(Dec), and the performance achieved when
scalar quantization with a subsequent index-reuse~(IR) is used for
encoding. 
Table entries labeled as N.A. (not available) indicate that those instances could not be considered here due to their high computational demand.\footnote{This instance would require a high-rate quantizer with a resolution larger than $L=16$.}
Notice that only the index assignments yielding best
possible performance were chosen for the experiments (e.g. a rate
of $R = 1$ [bits/sample] may be obtained from quantizers of resolution 
$L=4,8,16$).

Our simulation results reveal that simple index assignment
techniques applied to local clusters can achieve significant
performance gains using our coding approach,
especially for low data rates and strongly correlated sources.

\subsection{The CEO Problem}\label{s:CEO}
\noindent
In the following, we show the applicability of our techniques to another relevant sensor network model:
the quadratic Gaussian CEO Problem~\cite{Berger:CEO}.
Let $u_0$ be the output of a continuous-valued Gaussian source $U_0$. 
For all $n\in\mathcal{N}$, $\mathcal{N}=\{1,2,\sldots,N\}$, let $u_n$ denote noisy observations of $u_0$ which are corrupted by additive noise, i.e. $u_n=u_0+n_n$. The noise samples are generated by Gaussian noise processes $N_n$ statistically independent over $n$.
The observations $u_n$ are encoded and transmitted by independently operating encoders indexed by $n$.
The main task of the CEO is to estimate $u_0$ based on the data obtained from the encoders.  
In \cite{maierbacherB:CEOCoding} we derived the optimal decoding rule exploiting the special properties of this problem setup and studied a feasible decoder using a source approximation based on the factorization 
\begin{equation}
p(u_0,u_1,...,u_N)=p(u_0)\prod_{n =1}^N p(u_n|u_0)
\nonumber
\end{equation}
which can be easily represented by a factor graph~\cite{kschischang:factorgraphs}. 
In the following, we consider a scenario of $N=100$ encoders. 
The source process is Gaussian distributed $\mathcal{N}(s_0,\sigma_0^2)$ with mean $s_0=0$ and variance $\sigma_0^2=1$. 
The noise processes are Gaussian distributed $\mathcal{N}(l_n,\lambda_n^2)$ with mean $l_n=0$ and variance  $\lambda_n^2 = \lambda$ for all $n\in\mathcal{N}$ where $\lambda = \{0.1, 0.5\}$ was chosen depending on the considered scenario.
All scalar quantizers at the encoders are Lloyd-Max optimized to minimize the MSE in the sensor readings $u_n$ using identical resolution for quantization and identical rates for data transmission, i.e. $|\mathcal{I}_n| = L$ and $R_n = R$ for all $n \in \mathcal{N}$ where $L \leq 16$ was chosen. 
We use the scalable decoder as described in Section~\ref{s:setup} where the required PMFs 
were determined using Monte Carlo simulation with resolution $|\mathcal{I}_0|=64$ for the source $u_0$ and $|\mathcal{I}_n|=L$ for the observations $u_n$ for all $n\in\mathcal{N}$. 
Notice that in case of our highly symmetric scenario, with $|\mathcal{I}_n|=L$, $\lambda_n^2=\lambda^2$ and $l_n=0$, the probabilities $p(i_n|i_0)$ can be considered identical for all $n \in \mathcal{N}$. 
Therefore, the index assignments need to be designed only once for a single, arbitrarily chosen cluster $\Lambda \subseteq \mathcal{N}$ with $|\Lambda|=S$ where $S=4$ was chosen.
After employing the IR algorithm described in Section~\ref{s:index} with $\Omega = \Lambda$ and $\Psi=\{0\}$,
the resulting index assignments can be assigned repeatedly to all clusters within the system. 
To evaluate the performance of our coding strategies, we measure the output SNR for 
$U_0$ given by
\begin{equation}
\textrm{Output SNR} = 10 \cdot \log_{10}\left(\frac{u_0^2}{(u_0-\hat{u}_0)^2}\right) \textrm{in dB}
\nonumber
\end{equation}
versus the (symmetric) encoder transmission rate averaged over $(N+1) \times 10000$ 
source samples and compare it with the (sum) rate-distortion function, offered by
\cite{Berger:CEORD}, which presents an upper bound found to be tight for noise processes
with identical variance.
In Table \ref{tab:resultsCEO} we present some results to underline the effectiveness 
of our approach. 
The performance of the system without index assignments~(Dec) and the performance 
obtained by using index-reuse~(IR) is compared to the theoretically possible value as given by the (sum) rate-distortion function~(R/D) according to~\cite{Berger:CEORD}. 
Table entries labeled as N.A. (not available) indicate that those instance could not be considered here due to their high computational demand.\footnote{This instance would require a high-rate quantizer with a resolution larger than $L=16$.}
Table entries labeled as N.B. (no benefit) indicate that in this case index-reuse does not outperform standard quantization. 
Notice that only the index assignments yielding the best possible performance were chosen for the experiments.
\begin{table}
\caption{Simulation results for the CEO scenario with $N=100$ encoders, 
source variance $\sigma_0^2$=1 and noise-variances $\lambda^2$=\{0.1, 0.5\}.}
\begin{center}
\begin{tabular}{|l|c|c|c|c|c|c|c|c|}
\hline
					&\multicolumn{4}{|c|}{$\sigma_0^2$=1, $\lambda^2$=0.1}		&
					\multicolumn{4}{|c|}{$\sigma_0^2$=1, $\lambda^2$=0.5}		\\
$R$ [bit]				&1	&2	&3	&4	&
					1	&2	&3	&4	\\
\hline
$\textrm{SNR}_{\textrm{R/D}}$[dB]	&28.66	&29.70	&29.93	&29.99	&
					21.72	&22.74	&22.96	&23.01	\\
\hline
$\textrm{SNR}_{\textrm{Dec}}$[dB]	&9.67	&19.37	&26.21	&28.48	&
					15.25	&20.84	&22.49	&22.74	\\
\hline
$\textrm{SNR}_{\textrm{IR}}$[dB]	&22.71	&26.70	&28.21	&N.A.	&
					18.76	&21.56	&N.B.	&N.A.	\\
\hline
\end{tabular}
\end{center}
\label{tab:resultsCEO}
\vspace{-0.4cm}
\end{table}

The numerical results reveal that our index-reuse approach leads in many cases to significant performance improvements over standard quantization. 
It might happen, however, that our index assignments are not able to outperform scalar quantization.
Whether or not this is true depends on several factors: (a)~the quantizer resolution $L$, (b)~the number of output bits $R$ and (c)~the correlation properties of the sources determined by $\sigma_0^2$ and $\lambda$. In cases where the sources are weakly correlated, e.g. for large values of $\lambda$, it becomes harder (or even impossible) to find index assignments that offer good rate/distortion trade-offs. In particular this might be true in our case due to the simplicity of the considered coding concept and the sub-optimality of the proposed index-reuse algorithm whose performance is highly dependent on the choice of $L$ and $R$. 
      
\section{Concluding Remarks}\label{s:conclusion}
\noindent
We presented a scalable solution for distributed source coding in large-scale sensor networks. 
Our methods rely on the combination of a simple encoding stage (a scalar quantizer and an index assignment stage) and a source-optimized clustering algorithm. 
Despite the simplicity of the proposed techniques, our results show significant performance gains in comparison
with standard scalar quantization.
It is worth mentioning that the same ideas can be used together with other distributed source coding schemes, e.g. those based on syndromes~\cite{sandeep-kannan:discus}, on Diophantine index assignments~\cite{maierbacherB:dio} or even on random index assignments.
As part of our ongoing work we are considering the case in which the covariance matrix of the sensor observations is not known beforehand. Thus, each sensor must decide on-the-fly which code to use and inform the decoder. Finding distributed clustering and coding algorithms for this problem remains a challenging task.

\appendices
\noindent

\section{Optimal Decoding Rule}
\label{a:optDec}
\noindent
In this section, we want to derive a simple expression for the optimal decoding rule.

Let $k \in \mathcal{N}$ be the index identifying the source for which the estimate has to be calculated. 
Let $\mathcal{T} \subseteq \mathcal{N}$ be a set of indices identifying the encoders whose codewords, collected in the vector $\mathbf{w}_{\mathcal{T}} \in \mathcal{W}_{\mathcal{T}} = \prod_{\forall n \in \mathcal{T}} \mathcal{W}_n$, are available for the calculation.

Specifically, we want to show that $E\{U_k|\mathbf{w}_{\mathcal{T}}\} = \sum_{i_k = 0}^{|\mathcal{I}_k|-1} E\{U_k|i_k\} \cdot p(i_k|\mathbf{w}_{\mathcal{T}})$. We start by using the definition of the conditional expectation and apply the Bayes rule such that
\begin{equation}
E\{U_k|\mathbf{w}_{\mathcal{T}}\} = \int_{u_k = -\infty}^{+\infty} u_k \cdot p(u_k|\mathbf{w}_{\mathcal{T}})~d u_k = \frac{1}{p(\mathbf{w}_{\mathcal{T}})} \cdot \int_{u_k = -\infty}^{+\infty} u_k \cdot p(\mathbf{w}_{\mathcal{T}}|u_k) \cdot p(u_k)~d u_k.
\label{eqn:app00}
\end{equation}
Furthermore, we can state that
\begin{equation}
p(\mathbf{w}_{\mathcal{T}}|u_k) = \sum_{i_k = 0}^{|\mathcal{I}_k|-1} p(\mathbf{w}_{\mathcal{T}}|i_k) \cdot p(i_k|u_k) = 
\begin{cases} 
  p(\mathbf{w}_{\mathcal{T}}|i_k),  	& \mbox{if } q_k(u_k) = i_k \\
  0, 					& \mbox{otherwise}.
\end{cases} 
\nonumber
\end{equation}
Since the index $i_k=q_k(u_k)$ is constant
for all $u_k$ that fall into the quantizer region $\mathcal{B}_k(i_k)$ such that $b_k(i_k) < u_k \leq b_k(i_k+1)$, the integral in~(\ref{eqn:app00}) can be splitted into separate parts and we obtain 
\begin{equation}
E\{U_k|\mathbf{w}_{\mathcal{T}}\} 	
 			=	\frac{1}{p(\mathbf{w}_{\mathcal{T}})} \cdot \sum_{i_k = 0}^{|\mathcal{I}_k|-1} p(\mathbf{w}_{\mathcal{T}}|i_k) \int_{u_k = b_k(i_k)}^{b_k(i_k + 1)} u_k  \cdot p(u_k)~d u_k. 
\label{eqn:app1}
\end{equation}
We observe that 
$p(\mathbf{w}_{\mathcal{T}}|i_k) = \frac{p(i_k|\mathbf{w}_{\mathcal{T}}) \cdot p(\mathbf{w}_{\mathcal{T}})}{p(i_k)}$
and that
\begin{equation}
\frac{1}{p(i_k)} \cdot \int_{u_k = b_k(i_k)}^{b_k(i_k + 1)} u_k  \cdot p(u_k)~d u_k 
\overset{\mathrm{(a)}}{=} \int_{u_k = -\infty}^{+\infty} u_k  \cdot p(u_k|i_k)~d u_k 
= E\{U_k|i_k\}, 
\end{equation}
where the equality~(a) holds since $p(i_k|u_k)$ is either zero or unity depending on the fact if $q_k(u_k) = i_k$ or, identically, if $u_k$ falls into the quantizer region $\mathcal{B}_k(i_k)$ such that $b_k(i_k) < u_k \leq b_k(i_k+1)$.
Therefore, we can state that
\begin{equation}
p(u_k|i_k) = \frac{p(u_k,i_k)}{p(i_k)} = \frac{p(i_k|u_k) \cdot p(u_k)}{p(i_k)} =
\begin{cases}
  \frac{p(u_k)}{p(i_k)},  	& \mbox{if } q_k(u_k) = i_k \\
  0, 				& \mbox{otherwise}.
\end{cases}
\label{eqn:condProbA}
\end{equation}
Using these results together with~(\ref{eqn:app1}), the desired equality can be established easily.
\section{Efficient Marginalization and its Complexity}
\label{a:marg}
\noindent
In this section, we want to characterize the complexity of the marginalization operation required at several points of this work, e.g. consider the calculation of the optimal estimate in~(\ref{eq:estimatesOpt}) where the marginalization in~(\ref{eq:prob}) has to be performed using the argument in~(\ref{eq:prob2}). 

For a general treatment of the problem, we shall employ the same definitions as provided in Appendix~\ref{a:optDec}. Let furthermore $\mathcal{S} = \{k\} \cup \mathcal{T}$ be a set of indices identifying the sources whose discrete representations, collected in the vector $\mathbf{i}_{\mathcal{S}} \in \mathcal{I}_{\mathcal{S}} = \prod_{\forall n \in \mathcal{S}} \mathcal{I}_n$, are involved within the calculation. 
Specifically, we shall consider the calculation of $p(i_k = l|\mathbf{w}_{\mathcal{T}})$ out of $p(\mathbf{w}_{\mathcal{T}},\mathbf{i}_{\mathcal{S}})$ through the following marginalization
\begin{equation}
p(i_k = l| \mathbf{w}_{\mathcal{T}}) = \gamma \cdot \sum_{\forall \mathbf{i}_{\mathcal{S}} \in \mathcal{I}_{\mathcal{S}} : i_k = l} p(\mathbf{w}_{\mathcal{T}},\mathbf{i}_{\mathcal{S}})
\label{eqn:marg1}
\end{equation}
with $\gamma = 1\slash p(\mathbf{w}_{\mathcal{T}})$ and 
\begin{equation}
p(\mathbf{w}_{\mathcal{T}},\mathbf{i}_{\mathcal{S}}) = p(\mathbf{i}_{\mathcal{S}}) \cdot p(\mathbf{w}_{\mathcal{T}}|\mathbf{i}_{\mathcal{S}}) \overset{\mathrm{(a)}}{=} p(\mathbf{i}_{\mathcal{S}}) \cdot p(\mathbf{w}_{\mathcal{T}}|\mathbf{i}_{\mathcal{T}}) = p(\mathbf{i}_{\mathcal{S}}) \cdot \prod_{\forall n \in \mathcal{T}} p(w_n|i_n),
\label{eqn:marg2}
\end{equation} 
where equality~(a) obviously holds for $k \in \mathcal{T}$, since in this case $\mathcal{S} = \mathcal{T}$, and also for $k \notin \mathcal{T}$, since in this case $i_k$ does not provide any information about $\mathbf{w}_{\mathcal{T}}$ due to the fact that $\mathbf{i}_{\mathcal{T}}$ is known and $w_n = m_n(i_n)$ for all $n \in \mathcal{T}$ such that $p(\mathbf{w}_{\mathcal{T}}|\mathbf{i}_{\mathcal{S}}) = p(\mathbf{w}_{\mathcal{T}}|\mathbf{i}_{\mathcal{T}})$.

In the most straightforward implementation of the marginalization in~(\ref{eqn:marg1}), the summation over $p(\mathbf{w}_{\mathcal{T}},\mathbf{i}_{\mathcal{S}})$ has to be performed over all possible realizations of $\mathbf{i}_{\mathcal{S}} \in \mathcal{I}_{\mathcal{S}}$ with $i_k = l$ where the actual value of $p(\mathbf{w}_{\mathcal{T}},\mathbf{i}_{\mathcal{S}})$ can be calculated using the product representation in~(\ref{eqn:marg2}). It is worth pointing out that $p(\mathbf{w}_{\mathcal{T}}|\mathbf{i}_{\mathcal{T}})$  in~(\ref{eqn:marg2}) can become either zero or unity depending on the current configuration of the transition probabilities $p(w_n|i_n)$ for all $n \in \mathcal{T}$. This can be used to restrict the number of index tuples $\mathbf{i}_{\mathcal{S}} \in \mathcal{I}_{\mathcal{S}}$ that have to be considered throughout the marginalization in~(\ref{eqn:marg1}), as shown in the following.
%
%

For brevity, we shall restrict ourselves to the case where $k \in \mathcal{T}$, i.e. where $\mathcal{S}=\mathcal{T}$.\footnote{The results for the case $k \notin \mathcal{T}$ can be derived accordingly.} Let $\mathcal{Q}_{\mathcal{T}}(\mathbf{w}_{\mathcal{T}})$ be the set of index tuples $\mathbf{i}_{\mathcal{T}} \in \mathcal{I}_{\mathcal{T}}$ that are mapped onto $\mathbf{w}_{\mathcal{T}} \in \mathcal{W}_{\mathcal{T}}$.\footnote{It is worth pointing out that $\mathcal{Q}_{\mathcal{T}}(\mathbf{w}_{\mathcal{T}})$ can be constructed easily since the mapping functions $m_n$ are assumed to be known for all $n \in \mathcal{T}$.}
Then, the marginalization in~(\ref{eqn:marg1}) can be expressed as follows:
\begin{eqnarray}
p(i_k = l| \mathbf{w}_{\mathcal{T}}) 
& = & \gamma \cdot \sum_{\forall \mathbf{i}_{\mathcal{T}} \in \mathcal{Q}_{\mathcal{T}}(\mathbf{w}_{\mathcal{T}}) : i_k = l} p(\mathbf{i}_{\mathcal{T}}) \nonumber \\
& = & 
\begin{cases}
\gamma \cdot \displaystyle \sum_{\forall \mathbf{i}_{\mathcal{T}} \in \{i_k = l\} \times \mathcal{Q}_{\mathcal{T}\backslash\{k\}}(\mathbf{w}_{\mathcal{T}\backslash\{k\}})} p(\mathbf{i}_{\mathcal{T}}), & \textrm{if}~m_k(i_k=l) = w_k \\
0, & \textrm{otherwise}.
\end{cases}
\label{eqn:margX}
\end{eqnarray}
Notice that the marginalization according to~(\ref{eqn:margX}) has to be performed, if it has to be performed at all, only over the members of the set $\mathcal{Q}_{\mathcal{T}\backslash\{k\}}(\mathbf{w}_{\mathcal{T}\backslash\{k\}})$. Since the cardinality of this set is much smaller than the cardinality of $\mathcal{I}_{\mathcal{T}}$ in~(\ref{eqn:marg1}) the complexity of the marginalization can be reduced considerably. 

For a more detailed discussion of the complexity, the cardinality of the set $\mathcal{Q}_{\mathcal{T}}(\mathbf{w}_{\mathcal{T}})$ shall be characterized in the following. Notice that $\mathcal{Q}_n(w_n)$ denotes the set of indices $i_n \in \mathcal{I}_n$ that are mapped onto the codeword $w_n \in \mathcal{W}_n$. 
The following result is usefull
\begin{lemma}
\label{lem:maxIndex}
For any surjective mapping function $m_n: ~~~ \mathcal{I}_n \rightarrow \mathcal{W}_n$ and any $w_n \in \mathcal{W}_n$, 
$|\mathcal{Q}_n(w_n)| \leq |\mathcal{I}_n| - |\mathcal{W}_n| + 1$.
\end{lemma}
\begin{proof}
Since $m_n$ is a function each $i_n \in \mathcal{I}_n$ is mapped to exactly one $w_n \in \mathcal{W}_n$. From this we conclude that
(a) there are no $i_n \in \mathcal{I}_n$ that are mapped to more than one $w_n \in \mathcal{W}_n \Rightarrow \bigcap_{\forall w_n \in \mathcal{W}_n} \mathcal{Q}_n(w_n) = \emptyset$ (mutual exclusivity) $\Rightarrow |\bigcup_{\forall w_n \in \mathcal{W}_n} \mathcal{Q}_n(w_n)| = \sum_{\forall w_n \in \mathcal{W}_n} |\mathcal{Q}_n(w_n)|$ and
(b) each $i_n \in \mathcal{I}_n$ is mapped to some $w_n \in \mathcal{W}_n \Rightarrow |\bigcup_{\forall w_n \in \mathcal{W}_n} \mathcal{Q}_n(w_n)| = |\mathcal{I}_n|$.
Since $m_n$ is a surjective function there exists an $i_n \in \mathcal{I}_n$ for any $w_n \in \mathcal{W}_n$ such that $w_n = m_n(i_n)$ and we conclude that
(c) $|\mathcal{Q}_n(w_n)| \geq 1$ for all $w_n \in \mathcal{W}_n$.
From (a) and (b) we obtain that 
$\sum_{\forall w_n \in \mathcal{W}_n} |\mathcal{Q}_n(w_n)| = |\mathcal{I}_n|$
which can be solved for an arbitrarily chosen $w_n \in \mathcal{W}_n$, e.g. $w_n = a$, and we obtain
$|\mathcal{Q}_n(w_n = a)| = |\mathcal{I}_n| - \sum_{\forall w_n \in \mathcal{W}_n: w_n \neq a} |\mathcal{Q}_n(w_n)|$.
Because of (c) we know that $|\mathcal{Q}_n(w_n = a)|$ is maximal if $|\mathcal{Q}_n(w_n)| = 1$ for all $w_n \in \mathcal{W}_n: w_n \neq a$ and we obtain that $|\mathcal{Q}_n(w_n=a)| \leq |\mathcal{I}_n| - (|\mathcal{W}_n|-1)$ for any $a \in \mathcal{W}_n$ establishing the claim.
\end{proof}
Using this result, the complexity of the marginalization in~(\ref{eqn:marg1}) can be characterized. 
For the sake of a simple discussion, we assume that $|\mathcal{I}_n| = L$ for all $n \in \mathcal{T}$ and that $|\mathcal{W}_n| = K$ for all $n \in \mathcal{T}$. Using Lemma~\ref{lem:maxIndex} and after defining the system specific parameter $F=L-K+1$, we are able to conclude that in our case $|\mathcal{Q}_n(w_n)| \leq F$ for any $n \in \mathcal{T}$.
We furthermore assume that the elementary operations (like additions, multiplications, comparisons, look-ups, etc.) are of constant complexity, i.e. of $\mathcal{O}(1)$. Specifically, we assume that $p(\mathbf{i}_{\mathcal{T}})$ can be determined with a complexity of $\mathcal{O}(1)$, e.g. that it can be approximated, simulated, etc. with constant complexity or that it can be looked-up.

In the following, we consider the newly derived expression for the marginalization~(\ref{eqn:margX}). In the case where $m_k(i_k=l) = w_k$ holds, it is easy to see that, in the worst-case, $F^{|\mathcal{T}|-1}$ instances of $p(\mathbf{i}_{\mathcal{T}})$ are required throughout the calculation and that around $F^{|\mathcal{T}|-1}$ additions have to be performed. Thus, around $2 \cdot F^{|\mathcal{T}|-1}$ elementary operations have to be performed corresponding to a computational complexity of $\mathcal{O}(F^{|\mathcal{T}|-1})$. In the case where $m_k(i_k=l) = w_k$ does not hold, the result of the marginalization becomes zero, without any further calculations, and the computational complexity derives to be of $\mathcal{O}(1)$ for testing the case alone. 
%
%

\section{Complexity of Optimal Decoding}
\label{a:implOptDec}
\noindent
In this section, which uses the same definitions as the previous appendices, we discuss the complexity of optimal decoding as required e.g. in~(\ref{eq:estimatesOpt}).
Specifically, we want to consider the calculation 
\begin{equation}
\hat{u}_k(\mathbf{w}_{\mathcal{T}}) = \sum_{i_k = 0}^{|\mathcal{I}_k|-1} \tilde{u}_{k,i_k} \cdot p(i_k|\mathbf{w}_{\mathcal{T}}).
\label{eqn:estP}
\end{equation}
We observe that the calculation in~(\ref{eqn:estP}) requires that $\tilde{u}_{k,i_k}$ and $p(i_k|\mathbf{w}_{\mathcal{T}})$ have to be determined, multiplied and summed-up for all possible realizations of $i_k \in \mathcal{I}_k$ where $p(i_k|\mathbf{w}_{\mathcal{T}})$ can be derived from $p(\mathbf{i}_{\mathcal{S}})$ using the efficient marginalization described in Appendix~\ref{a:marg}.

In order to use the results derived in Appendix~\ref{a:marg}, we restrict ourselves to the case where $k \in \mathcal{T}$, i.e. where $\mathcal{S} = \mathcal{T}$.\footnote{The results for the case $k \notin \mathcal{T}$ can be derived accordingly.}
For a simplified complexity analysis, we furthermore assume that $|\mathcal{I}_n| = L$, $|\mathcal{W}_n| = K$ and $|\mathcal{Q}_n(w_n)| \leq F$ for all $n \in \mathcal{T}$ where  $F=L-K+1$. 
Elementary operations (like additions, multiplications, comparisons, look-ups, etc.) are assumed to be of constant complexity, i.e. of $\mathcal{O}(1)$. 
Specifically, we assume that $\tilde{u}_{k,i_k}$ and $p(\mathbf{i}_{\mathcal{T}})$ can be determined with a complexity of $\mathcal{O}(1)$, e.g. that they can be approximated, simulated, etc. with constant complexity or that they can be looked-up.

Using the results of Appendix~\ref{a:marg}, we are able to state that the computational complexity for deriving $p(i_k|\mathbf{w}_{\mathcal{T}})$, as required in~(\ref{eqn:estP}), is of $\mathcal{O}(F^{|\mathcal{T}|-1})$ or of $\mathcal{O}(1)$ depending on the fact whether $i_k$ is mapped onto $w_k$ or not. In order to determine the overall complexity, we notice that the summation in~(\ref{eqn:estP}) has to be performed over all $i_k \in \mathcal{I}_k$. Therefore, we can employ Lemma~\ref{lem:maxIndex} in Appendix~\ref{a:marg} to determine how often it will be true (at most) that $m_k(i_k) = w_k$ and, thus, how often (at most) the calculation of $p(i_k|\mathbf{w}_{\mathcal{T}})$ in~(\ref{eq:prob}) has to be performed. We conclude that this calculation has to be performed (at most) $F$ times. 
The test if $m_k(i_k) = w_k$ is true, the look-up of $\tilde{u}_{k,i_k}$ as well as the multiplication in~(\ref{eqn:estP}) can be neglected compared to the complexity of calculating $p(i_k|\mathbf{w}_{\mathcal{T}})$ in total $F$ times. Therefore, we are able to conclude that calculating one estimate has a computational complexity of $\mathcal{O}(F^{|\mathcal{T}|})$.
%

\section{Distortion Calculation}
\label{a:dist}
\noindent
In this section, which again uses the same definitions as the previous appendices, we shall show that the overall distortion associated with each source $k \in \mathcal{N}$ can be described by the sum
\begin{equation}
E\{(\hat{U}_k-U_k)^2\}=E\{(\tilde{U}_k-U_k)^2\}+E\{(\hat{U}_k-\tilde{U}_k)^2\}
\label{eqn:distP1}
\end{equation}
where $E\{(\tilde{U}_k-U_k)^2\}$ is the distortion caused by the quantization stage and $E\{(\hat{U}_k-\tilde{U}_k)^2\}$ is the distortion caused by the index assignment stage.

To do so, we start with the definition of the expectation value and obtain
\begin{equation}
E\{(\hat{U}_k - U_k)^2\} = \! \! \! \sum_{\forall{w}_{\mathcal{T}} \in \mathcal{W}_{\mathcal{T}}} \!  p(\mathbf{w}_{\mathcal{T}}) E\{(\hat{U}_k - U_k)^2|\mathbf{w}_{\mathcal{T}}\}
\overset{\mathrm{(a)}}{=} \! \! \! \sum_{\forall{i}_{\mathcal{T}} \in \mathcal{I}_{\mathcal{T}}} \!  p(\mathbf{i}_{\mathcal{T}}) E\{(\hat{U}_k - U_k)^2|\mathbf{i}_{\mathcal{T}}\} \nonumber
\end{equation}
where the equality~(a) holds due to the fact that the index assignments $m_n$ are surjective functions for all $n \in \mathcal{T}$ and, thus, the summation over all $\mathbf{w}_{\mathcal{T}} \in \mathcal{W}_{\mathcal{T}}$ covers the same observation space as the summation over all $\mathbf{i}_{\mathcal{T}} \in \mathcal{I}_{\mathcal{T}}$.
Based on this observation, equation~(\ref{eqn:distP1}) can easily be established by showing that 
\begin{equation}
E\{(\hat{U}_k - U_k)^2|\mathbf{i}_{\mathcal{T}}\}= E\{(\tilde{U}_k - U_k)^2|\mathbf{i}_{\mathcal{T}}\} + E\{(\hat{U}_k - \tilde{U}_k)^2|\mathbf{i}_{\mathcal{T}}\}.
\label{eqn:distP2}
\end{equation}
The definition of the conditional expectation allows us to rewrite
\begin{eqnarray}
E\{(\hat{U}_k - U_k)^2|\mathbf{i}_{\mathcal{T}}\} & \overset{\mathrm{(a)}}{=} & \int_{u_k = - \infty}^{+\infty} (\hat{u}_k(\mathbf{w}_{\mathcal{T}}) - u_k)^2 p(u_k|\mathbf{i}_{\mathcal{T}}) du_k \nonumber \\
& \overset{\mathrm{(b)}}{=} & \int_{u_k = - \infty}^{+\infty} (\hat{u}_k(\mathbf{w}_{\mathcal{T}}) - u_k)^2 p(u_k|i_k) du_k \nonumber \\
& \overset{\mathrm{(c)}}{=} & \frac{1}{p(i_k)} \int_{u_k = b_k(i_k)}^{b_k(i_k+1)} (\hat{u}_k(\mathbf{w}_{\mathcal{T}}) - u_k)^2 p(u_k) du_k
\label{eqn:cmDistA}
\end{eqnarray}
where the definition of the conditional expectation is used in~(a) together with the fact that $w_n=m_n(i_n)$ for all $n \in \mathcal{T}$, equality~(b) is valid since $i_k$ is known if $\mathbf{i}_{\mathcal{T}}$ is known and, thus, $p(u_k|\mathbf{i}_{\mathcal{T}}) = p(u_k|i_k)$ and equality~(c) holds due to~(\ref{eqn:condProbA}).
Assuming that 
\begin{equation}
\tilde{u}_{k,i_k} = \frac{\int_{u_k = b_k(i_k)}^{b_k(i_k+1)} u_k \cdot p(u_k) du_k}{\int_{u_k = b_k(i_k)}^{b_k(i_k+1)} p(u_k) du_k} = \frac{\int_{u_k = b_k(i_k)}^{b_k(i_k+1)} u_k \cdot p(u_k) du_k}{p(i_k)},
\end{equation}
i.e. that the reconstruction value of the quantizer is the centroid of the quantization region, it is possible to show that the required integration can be split into two parts\footnote{This can be achieved by substituting $\hat{u}_k(\mathbf{w}_{\mathcal{T}}) = \tilde{u}_{k,i_k} + d_k$, where $d_k = \hat{u}_k(\mathbf{w}_{\mathcal{T}}) - \tilde{u}_{k,i_k}$, such that $(\hat{u}_k(\mathbf{w}_{\mathcal{T}}) - u_k)^2$ can be expressed as $(\tilde{u}_{k,i_k} + d_k - u_k)^2$ which derives to $(\tilde{u}_{k,i_k}-u_k)^2 + 2d_k(\tilde{u}_{k,i_k} - u_k) + d_k^2$.} such that 
\begin{equation}
\begin{split}
\displaystyle \int_{u_k = b_k(i_k)}^{b_k(i_k+1)} (\hat{u}_k(\mathbf{w}_{\mathcal{T}}) - u_k)^2 p(u_k) du_k = 
\int_{u_k = b_k(i_k)}^{b_k(i_k+1)} (\tilde{u}_{k,i_k} - u_k)^2 p(u_k) du_k  \\ +
(\hat{u}_k(\mathbf{w}_{\mathcal{T}}) - \tilde{u}_{k,i_k})^2 p(\tilde{u}_{k,i_k}),
\end{split}
\label{eqn:sumIntA}
\end{equation}
where $p(\tilde{u}_{k,i_k}) = p(i_k)$.
Plugging~(\ref{eqn:sumIntA}) into~(\ref{eqn:cmDistA}), we obtain
\begin{eqnarray}
E\{(\hat{U}_k-U_k)^2|\mathbf{i}_{\mathcal{T}}\} & = & \frac{1}{p(i_k)} \int_{u_k = b_k(i_k)}^{b_k(i_k+1)} \! \! (\tilde{u}_{k,i_k} - u_k)^2 p(u_k) du_k + (\hat{u}_k(\mathbf{w}_{\mathcal{T}}) - \tilde{u}_{k,i_k})^2 \nonumber \\
	& = & E\{(\tilde{U}_k-U_k)^2|i_k\} + E\{(\hat{U}_k-\tilde{U}_k)^2|\mathbf{i}_{\mathcal{T}}\}
\end{eqnarray}
directly establishing~(\ref{eqn:distP2}) and, thus, the desired result in~(\ref{eqn:distP1}).

\section{Complexity of the Index-Reuse Optimization}
\label{a:complexityIR}
\noindent
The computational complexity of the index-reuse algorithm presented in Algorithm~\ref{alg:indexreuse} can be bounded by evaluating how often, in the worst-case, the operations within the innermost of the nested loops have to be performed. 

The outermost loop is executed for each of the $L-K$ mergings that have to be performed to obtain mappings with $K$ output codewords from the initial mapping with $L$ output codewords. The second loop is executed for all considered encoders $n \in \Omega$, i.e. in total $|\Omega|$ times. Finally, the innermost loop runs through all possibilities of choosing $2$ out of $k$ codewords for the merging, i.e. we have $\binom{k}{2} = \frac{k!}{2!(k-2)!} = \frac{1}{2}(k^2-k)$ possibilities. In the worst-case, i.e. in the initial case where $k=L$, we obtain $\frac{1}{2}(L^2-L)$ possibilities. Thus, the operations within the innermost loop have to be performed $|\Omega|(L-K)\frac{1}{2}(L^2-L) = \frac{1}{2}|\Omega|(L-K)L^2- \frac{1}{2}|\Omega|(L-K)L$ times in the worst-case.

Now, to determine the overall complexity of the algorithm the complexities of the merging operation $\mathbf{e}_n = g(\mathbf{f}_n,a,b)$, the complexity of the distortion calculation $d = d_d(\Psi,\mathcal{E}_n)$ and the test if $d<d^*$ have to be determined. Assuming that the merging and the test can be performed with a constant computational complexity of $\mathcal{O}(1)$, it remains to determine the complexity of calculating $d_d(\Psi)$ given the current set of mapping functions $\mathcal{E}_n$. 
This calculation requires the calculation of $d_d(n)$ according to~(\ref{eqn:distD}) which, in turn, requires the calculation of the estimate $\hat{u}_n(\mathbf{w}_{\Omega})$ according to~(\ref{eq:estimatesOpt}) for all $L^{|\Psi|}$ possible realizations of $\mathbf{i}_{\Psi} \in \mathcal{I}_{\Psi}$.\footnote{There are more efficient ways to calculate $d_d(n)$ based on intermediate results. However, due to lack of space, the discussion is neglected here.}
Assuming that $n \in \Omega$, i.e. that $|\Psi|=|\Omega|$, the result of Appendix~\ref{a:implOptDec} directly applies here and we can state that calculating one estimate has a computational complexity of $\mathcal{O}(F^{|\Omega|})$.\footnote{The results for the case where $n \notin \Omega$ can be derived accordingly.} Thus, in the most straightforward implementation, the computational complexity of calculating $d_d(n)$ according to~(\ref{eqn:distD}) is of $\mathcal{O}((LF)^{|\Omega|})$.

Based on the presented results and after substituting $F$ by $L-K+1$, as derived in Appendix~\ref{a:marg} for surjective mapping functions, we are able to conclude that the overall computational complexity of the index-reuse algorithm is of $\mathcal{O}(|\Omega| L^{|\Omega|+2} (L-K)^{|\Omega|+1})$ showing an exponential growth with $|\Omega|$.
\section{Efficient Sub-Optimal Decoding and its Complexity}
\label{a:subOptDec}
\noindent
In this section, which uses the same definitions as the previous appendices, we shall elaborate on how a source factorization based on CCREs
according to~(\ref{eqn:CCREfactorization}) and~(\ref{eqn:CCREConditions}) can be used for efficient decoding. 
This shall be achieved by assuming that the factorization~(\ref{eqn:CCREfactorization}) also holds for the discrete case\footnote{This assumption is plausible since $U_n \rightarrow I_n$ forms a Markov chain for $n = 1,2,\sldots,N$.} such that 
\begin{equation}
\hat{p}(\mathbf{i}) = \prod_{m = 1}^M f_m(\mathbf{i}_{\mathcal{S}_m}) =
\prod_{m = 1}^{M}p(\mathbf{i}_{\mathcal{A}_m}|\mathbf{i}_{\mathcal{B}_m}),
\label{eqn:af0}
\end{equation}
a fact, which can be exploited for scalable decoding as shown in~\cite{bar05} for a similar system setup. 
Since the decoder design considered in this work follows the same principles, we shall focus on the differences resulting from system specific properties.

In Section~\ref{s:setup} we have shown that the calculation of the optimal estimate $\hat{u}_n(\mathbf{w})$ according to~(\ref{eq:estimatesOpt}) for $n \in \mathcal{N}$ requires the calculation of the probabilities 
\begin{equation}
p(i_n = l|\mathbf{w}) = \gamma \cdot \sum_{\forall \mathbf{i} \in \mathcal{I}:i_n = l} p(\mathbf{w},\mathbf{i}) \overset{\mathrm{(a)}}{=} \gamma \cdot \sum_{\forall \mathbf{i} \in \mathcal{Q}(\mathbf{w}):i_n = l} p(\mathbf{i})
\label{eqn:af1}
\end{equation}
where the equality~(a) is due to the result derived in Appendix~\ref{a:marg}.
Replacing $p(\mathbf{i})$ by its approximation $\hat{p}(\mathbf{i})$ as given by the factorization in~(\ref{eqn:af0}), we obtain the following approximation 
\begin{equation}
\hat{p}(i_n = l|\mathbf{w}) = \gamma \cdot \sum_{\forall \mathbf{i} \in \mathcal{Q}(\mathbf{w}):i_n = l} \hat{p}(\mathbf{i}) = \gamma \cdot \sum_{\forall \mathbf{i} \in \mathcal{Q}(\mathbf{w}):i_n = l} \left ( \prod_{m=1}^{M}f_m(\mathbf{i}_{\mathcal{S}_m}) \right )
\label{eqn:af2}
\end{equation}
which can be calculated efficiently for all $l \in \mathcal{I}_n$ and for all $n \in \mathcal{N}$ by running the {\it sum-product algorithm} on the {\it factor graph} representation of the factorization in~(\ref{eqn:af0}). For a general treatment of factor graphs and the sum-product algorithm please refer to~\cite{kschischang:factorgraphs} or to~\cite{bar05} where a similar system setup is discussed. 

In order to provide the fundamentals, we include a brief review here. A factor graph is a bipartite graph that consists of {\it variable} and {\it function nodes} and expresses how a (global) function factors into (local) functions. The variable nodes represent the arguments of the functions and the function nodes the (local) functions itself. 
The sum-product algorithm allows us to perform the (global) marginalization in~(\ref{eqn:af2}) 
based on (local) marginalizations of the following type
\begin{equation}
\pmb\mu_{m \rightarrow n}(l) = \sum_{\forall \mathbf{i}_{\mathcal{S}_m} \in \mathcal{Q}_{\mathcal{S}_m}(\mathbf{w}_{\mathcal{S}_m}):i_n = l} \left(f_m(\mathbf{i}_{\mathcal{S}_m}) \prod_{g \in \mathcal{S}_m : g \neq n} \pmb\mu_{g \rightarrow m}\right)
\label{eqn:af3}
\end{equation}
which are performed in a structured way for all $n \in \mathcal{S}_m$ and $m \in \mathcal{M}$.
Following the intuition in~\cite{kschischang:factorgraphs}, the results of the marginalizations in~(\ref{eqn:af3}) for $l = 0,1,\sldots,|\mathcal{I}_n|-1$ can be seen as {\it messages} represented a vector $\pmb\mu_{m \rightarrow n}= (\mu_{m \rightarrow n}(0), \mu_{m \rightarrow n}(1), \sldots, \mu_{m \rightarrow n}(|\mathcal{I}_n|-1))$ that are {\it sent} from the function node $m \in \mathcal{M}$ to the variable node $n \in \mathcal{N}$ for further processing. 
Similarly, the inputs of the marginalizations in~(\ref{eqn:af3}) can also be seen as messages $\pmb\mu_{g \rightarrow m}$ that were {\it received} at the function node $m$ originating from some variable nodes $g \in \mathcal{N}$. Those messages represent the product 
\begin{equation}
\pmb\mu_{g \rightarrow m}(k) = \prod_{\forall h \in \mathcal{M}: g \in \mathcal{S}_h, h \neq m} \pmb\mu_{h \rightarrow g}(k)
\label{eqn:af4}
\end{equation}
for $k=0,1,\sldots,|\mathcal{I}_g|-1$ and, thus, $\pmb\mu_{g \rightarrow m} = (\mu_{g \rightarrow m}(0),\mu_{g \rightarrow m}(1),\sldots,\mu_{g \rightarrow m}(|\mathcal{I}_g|-1))$.
Using this abstraction, the techniques described in~\cite{kschischang:factorgraphs} can be directly applied here giving rise to an efficient calculation of~(\ref{eqn:af2}).\footnote{It is worth pointing out that the expression in~(\ref{eqn:af3}) is optimized to minimize the marginalization complexity by using knowledge about the received codewords. This in turn means that in this particular setup the function nodes have to be initialized and not the variable nodes as in conventional implementations. Specifically, the function nodes are initialized by defining the set $\mathcal{Q}_{\mathcal{S}_m}(\mathbf{w}_{\mathcal{S}_m})$ using knowledge of $\mathbf{w}_{\mathcal{S}_m}$ for $m = 1,2,\sldots,M$ and the variable nodes are initialized with trivial messages.}
%
In particular this is achieved by running an appropriate {\it message passing} algorithm\footnote{For factor graphs without cycles the efficient {\it forward-backward} algorithm can be employed, see~\cite{kschischang:factorgraphs}.} along the factor graph representation of~(\ref{eqn:af0}) and
depending on the fact if the message passing procedure terminates or not, i.e. if the factor graph is cycle-free or not,\footnote{Using the result in~\cite{bar05} this can be ensured if $|\mathcal{B}_m| = 1$ for $m=1,2,\sldots,M$.} the exact or an approximated value of $\hat{p}(i_n=l|\mathbf{w})$ is obtained simultaneously for all $l \in \mathcal{I}_n$ and $n \in \mathcal{N}$. 

Since the presented decoding scheme is based on message passing, its overall complexity can be analyzed by 
considering all messages that are created during the decoding process and jointly evaluating their complexity. 
We notice that the calculation of the messages at each function node $m \in \mathcal{M}$ according to~(\ref{eqn:af3}) requires a marginalization of the same type as discussed in Appendix~\ref{a:marg}. Assuming that furthermore $L_n = L$ and $K_n = K$ for $n = 1,2,\sldots,N$ and that the complexity of elementary operations are the same as stated in Appendix~\ref{a:marg}, the derived results directly apply here and we are able to conclude that the messages at the function nodes $m \in \mathcal{M}$ can be created with a computational complexity of $\mathcal{O}(F^{|\mathcal{S}_m|})$.
Considering the messages created at the variable nodes $n \in \mathcal{N}$ according to~(\ref{eqn:af4}), it is easy to see that the messages can be derived with a computational complexity of $\mathcal{O}(L)$. We notice that the complexity of calculating the messages at the function nodes is higher than at the variable nodes since generally $|\mathcal{S}_m| > 1$ for all (but maybe one) $m \in \mathcal{M}$, i.e. the complexity of calculating the messages at the variable nodes can be neglected here.
In order to provide an expression for the complexity, we have to distinguish between two cases.

In the case where the factor graph is cycle-free, the efficient {\it forward-backward} algorithm, see~\cite{kschischang:factorgraphs}, can be used for message passing and only one message (in each direction) needs to be passed along each edge within the graph. Assuming that $|\mathcal{S}_m| \leq S$ for all $m\in\mathcal{M}$, the calculation in~(\ref{eqn:af3}) has to be performed at most $M \cdot S$ times leading to a computational complexity of $\mathcal{O}(MSF^S)$.
In the case where the graph has cycles, the message passing has to be performed in an iterative way for an reasonable amount of iterations $T >> 1$, see~\cite{kschischang:factorgraphs}, and we obtain a computational complexity of $\mathcal{O}(TMSF^S)$.

\section{Complexity of Source-Optimized Clustering}
\label{a:cluster}
\noindent
The complexity of source-optimized clustering used in Section~\ref{s:clusteringAlgorithm} shall be discussed next.\footnote{The main goal at this point is to show that source-optimized clustering is of polynominal complexity (considering the number of sources $N$) and not to find an exact expression for the complexity of Algorithm~\ref{alg:clustering}. This would exceed the scope of this work.} In~\cite{jain99:clustering} it is shown that hierarchical clustering, upon which the presented procedure is based, has a computational complexity of $\mathcal{O}(N^2\log N)$.
However, these results do not directly hold for source-optimized clustering since for each step of the procedure, i.e. for each merging performed, the differential KLD benefit $\Delta D'(\Lambda_k',\Lambda_l')$ according to~(\ref{eqn:diffKLD}) has to be calculated. Looking at~(\ref{eqn:diffKLD}) in more detail, we observe that merging cluster $\Lambda_k'$ and $\Lambda_l'$ requires the calculation of the KLD benefit $\Delta D(\Lambda_k'\cup\Lambda_l',\emptyset)$ according to~(\ref{eqn:deltaKLD}) which, in turn, requires the calculation of the determinant for the corresponding covariance matrix $\mat{R}_{\Lambda_k' \cup \Lambda_l'}$. 
Using the general definition of determinants, it is easy to see that it can be calculated using Gaussian elimination. Assuming that the matrix, whose determinant has to be derived, is of size $N \times N$, then the complexity of the Gaussian elimination and, thus, also of calculating the determinant is of $\mathcal{O}(N^3)$, see e.g.~\cite{strassen:gaussianElimination}.\footnote{There are more efficient ways to calculate the determinant of a matrix, see e.g. \cite{strassen:gaussianElimination}, but it is sufficient for our purposes to assume that the Gaussian elimination is used.} Since $|\Lambda_k'\cup \Lambda_l'|$ is always smaller or equal to $N$, the matrix $\mat{R}_{\Lambda_k' \cup \Lambda_l'}$ is (at most) of size $N \times N$ and, thus, the calculation of $|\mat{R}_{\Lambda_k' \cup \Lambda_l'}|$ is of $\mathcal{O}(N^3)$. Assuming that the complexity of performing one merging step within the classical hierarchical clustering algorithm is of $\mathcal{O}(1)$, i.e. the minimum possible, then the number of mergings to be performed can be bounded by $\mathcal{O}(N^2\log N)$. For the source-optimized clustering procedure this means that its computational complexity is of $\mathcal{O}(N^5\log N)$. 

\section{Complexity of Source-Optimized Linking}
\label{a:link}
\noindent
This section addresses the complexity of constructing the source-optimized factorization presented in Section~\ref{s:clusterTrees}. Using the result in~\cite{nepomniaschaya:efficentEdmonds}, we are able to conclude that the directed spanning tree algorithm, upon which the presented procedure is based on, can be implemented with a complexity of $\mathcal{O}(C\log C)$. 
However, beside this, also the complexity of preprocessing the data required to initialize the directed spanning tree algorithm has to be considered. In particular this means that the link costs have to be determined before the directed spanning tree algorithm can be employed. 
In total there are $C^2$ link costs $c_{k,l}$ representing the KLD benefit associated with establishing a link between cluster $\Lambda_k$, $k \in \Gamma$, and $\Lambda_l$, $l \in \Gamma$, that have to be calculated according to~(\ref{eqn:cost}). In order to calculate this link cost all possible combinations of $\mathcal{P}_k' \in \mathcal{T}(A,\Lambda_k)$ and $\mathcal{Q}_l' \in \mathcal{T}(B,\Lambda_l)$ have to be evaluated as stated in~(\ref{eqn:linkSet}). It is easy to see that the number of such combinations is given by the product between $|\mathcal{T}(A,\Lambda_k)|$ and $|\mathcal{T}(B,\Lambda_l)|$ where $|\mathcal{T}(A,\Lambda_k)| = \binom{|\Lambda_k|}{A}$ and $|\mathcal{T}(B,\Lambda_l)| = \binom{|\Lambda_l|}{B}$. For simplicity, we assume in the following that $A = B = \frac{S}{2}$ and that $|\Lambda_c| = S$ for $c=1,2,\sldots,C$.\footnote{For other configurations the following results can be derived accordingly.} After simple mathematical manipulation we are able to conclude that there are at most $2^{S\log_2 S}$ such combinations. It remains to derive the complexity of calculating the argument in~(\ref{eqn:linkSet}), i.e. the complexity of calculating $\Delta D^{\ast}(\mathcal{P}_k',\mathcal{Q}_l')$ according to~(\ref{eqn:KLDLink}), which is clearly determined by the complexity of calculating $\Delta D(\mathcal{P}_k' \cup \mathcal{Q}_l',\emptyset)$ according to~(\ref{eqn:deltaKLD}). Using the same arguments as in Appendix~\ref{a:cluster}, we can state that the complexity of calculating the determinant $|\mat{R}_{\mathcal{P}_k' \cup \mathcal{Q}_l'}|$ in~(\ref{eqn:deltaKLD}) is of $\mathcal{O}(|\mathcal{P}_k' \cup \mathcal{Q}_l'|^3)$, i.e. it is of $\mathcal{O}(S^3)$ using the same assumptions as before. Putting everything together, we are able to conclude that the calculation of all link cost is of $\mathcal{O}(C^2 2^{S\log_2 S} S^3) = \mathcal{O}(C^2 2^{(3 + S)\log_2 S})$.
The computational complexity of the overall source-optimized linking procedure is then given by the sum of the derived complexities, i.e. of the directed spanning tree algorithm and the link cost calculation. Since the complexity of the algorithm can be neglected here, we conclude that the complexity of source-optimized linking is of $\mathcal{O}(C^2 2^{(3 + S)\log_2 S})$ which is only feasible for small values of $S$.

\bibliographystyle{plain}
\bibliography{library}

\begin{thebibliography}{}

\bibitem[\protect\citeauthoryear{Aho and Ullman}{Aho and
  Ullman}{1997}]{AhoUllman:Foundations}
{\sc Aho, A.~V.} {\sc and} {\sc Ullman, J.~D.} 1997.
\newblock {\em Foundations of Computer Science, C Edition}.
\newblock W. H. Freeman \& Co., New York, NY, USA.

\bibitem[\protect\citeauthoryear{Barros and Tuechler}{Barros and
  Tuechler}{2006}]{bar05}
{\sc Barros, J.} {\sc and} {\sc Tuechler, M.} 2006.
\newblock Scalable decoding on factor trees: A practical solution for sensor
  networks.
\newblock {\em IEEE Transactions on Communications\/}~{\em 54}, 284--294.

\bibitem[\protect\citeauthoryear{Berger, Zhang, and Viswanathan}{Berger
  et~al\mbox{.}}{1996}]{Berger:CEO}
{\sc Berger, T.}, {\sc Zhang, Z.}, {\sc and} {\sc Viswanathan, H.} 1996.
\newblock The {CEO} problem.
\newblock {\em IEEE Trans. Inform. Theory\/}~{\em 42}, 887--902.

\bibitem[\protect\citeauthoryear{Cardinal and Assche}{Cardinal and
  Assche}{2002}]{CardinalAssche:EntropyQuant}
{\sc Cardinal, J.} {\sc and} {\sc Assche, G.~V.} 2002.
\newblock Joint entropy-constrained multiterminal quantization.
\newblock In {\em Proceedings of the International Symposium on Information
  Theory}. Lausanne, Switzerland.

\bibitem[\protect\citeauthoryear{Chen, Zhang, Berger, and Wicker}{Chen
  et~al\mbox{.}}{2004}]{Berger:CEORD}
{\sc Chen, J.}, {\sc Zhang, X.}, {\sc Berger, T.}, {\sc and} {\sc Wicker,
  S.~B.} May 2004.
\newblock An upper bound on the sum-rate distortion function and its
  corresponding rate allocation schemes for the {CEO} problem.
\newblock {\em Special Issue of JSAC, On Fundamental Performance of Wireless
  Sensor Networks\/}.

\bibitem[\protect\citeauthoryear{Chu and Liu}{Chu and Liu}{1965}]{ChuLiu:MDST}
{\sc Chu, Y.~J.} {\sc and} {\sc Liu, T.~H.} 1965.
\newblock On the shortest arborescence of a directed graph.
\newblock {\em Science Sinica\/}~{\em 14}, 1396--1400.

\bibitem[\protect\citeauthoryear{Cover and Thomas}{Cover and
  Thomas}{1991}]{cover-thomas:it-book}
{\sc Cover, T.~M.} {\sc and} {\sc Thomas, J.} 1991.
\newblock {\em {Elements of Information Theory}}.
\newblock John Wiley and Sons, Inc.

\bibitem[\protect\citeauthoryear{Dietrich and Newsam}{Dietrich and
  Newsam}{1997}]{dietrich97fast}
{\sc Dietrich, C.~R.} {\sc and} {\sc Newsam, G.~N.} 1997.
\newblock Fast and exact simulation of stationary {Gaussian} processes through
  circulant embedding of the covariance matrix.
\newblock {\em SIAM Journal on Scientific Computing\/}~{\em 18,\/}~4,
  1088--1107.

\bibitem[\protect\citeauthoryear{Edmonds}{Edmonds}{1967}]{Edmonds:MDST}
{\sc Edmonds, J.} 1967.
\newblock Optimum branchings.
\newblock {\em J. Research of the National Bureau of Standards\/}~{\em 71B},
  233--240.

\bibitem[\protect\citeauthoryear{Flynn and Gray}{Flynn and
  Gray}{1987}]{FlynnG:87}
{\sc Flynn, T.~J.} {\sc and} {\sc Gray, R.~M.} 1987.
\newblock Encoding of correlated observations.
\newblock {\em IEEE Trans. Inform. Theory\/}~{\em IT-33,\/}~6, 773--787.

\bibitem[\protect\citeauthoryear{Georgiadis}{Georgiadis}{2003}]{Georgiadis:MDS%
T}
{\sc Georgiadis, L.} 2003.
\newblock Arborescence optimization problems solvable by {E}dmonds' algorithm.
\newblock {\em Theor. Comput. Sci.\/}~{\em 301,\/}~1-3, 427--437.

\bibitem[\protect\citeauthoryear{Jain, Murty, and Flynn}{Jain
  et~al\mbox{.}}{1999}]{jain99:clustering}
{\sc Jain, A.~K.}, {\sc Murty, M.~N.}, {\sc and} {\sc Flynn, P.~J.} 1999.
\newblock Data clustering: a review.
\newblock {\em ACM Computing Surveys\/}~{\em 31,\/}~3, 264--323.

\bibitem[\protect\citeauthoryear{Jayant and Noll}{Jayant and
  Noll}{1984}]{jayant-noll:coding-book}
{\sc Jayant, N.} {\sc and} {\sc Noll, P.} 1984.
\newblock {\em {D}igital {C}oding of {W}aveforms}.
\newblock Prentice Hall.

\bibitem[\protect\citeauthoryear{Kschischang, Frey, and Loeliger}{Kschischang
  et~al\mbox{.}}{2001}]{kschischang:factorgraphs}
{\sc Kschischang, F.~R.}, {\sc Frey, B.}, {\sc and} {\sc Loeliger, H.-A.} 2001.
\newblock Factor graphs and the sum-product algorithm.
\newblock {\em IEEE Trans. Inform. Theory\/}~{\em 47,\/}~2, 498--519.

\bibitem[\protect\citeauthoryear{Maierbacher and Barros}{Maierbacher and
  Barros}{2005}]{maierbacherB:CEOCoding}
{\sc Maierbacher, G.} {\sc and} {\sc Barros, J.} 2005.
\newblock Low-complexity coding for the {CEO} problem with many encoders.
\newblock In {\em Twenty-sixt Syposium on Information Theory in the Benelux}.
  Brussels, Belgium.

\bibitem[\protect\citeauthoryear{Maierbacher and Barros}{Maierbacher and
  Barros}{2006a}]{maierbacherB:KLD}
{\sc Maierbacher, G.} {\sc and} {\sc Barros, J.} 2006a.
\newblock On the {K}ullback-{L}eibler distance and the mean square distortion
  of mismatched distributed quantizers.
\newblock In {\em Proceedings of the 3rd International Workshop on Mathematical
  Techniques and Problems in Telecommunications}. Leiria, Portugal.

\bibitem[\protect\citeauthoryear{Maierbacher and Barros}{Maierbacher and
  Barros}{2006b}]{maierbacherB:Clustering}
{\sc Maierbacher, G.} {\sc and} {\sc Barros, J.} 2006b.
\newblock Source-optimized clustering for distributed source coding.
\newblock In {\em Proceedings of the IEEE Global Telecommunications Conference
  (GLOBECOM'06)}. San Francisco, USA.

\bibitem[\protect\citeauthoryear{Maierbacher and Barros}{Maierbacher and
  Barros}{2007}]{maierbacherB:dio}
{\sc Maierbacher, G.} {\sc and} {\sc Barros, J.} 2007.
\newblock Diophantine index assignments for distributed source coding.
\newblock In {\em Proceedings of the 2007 IEEE Information Theory Workshop (ITW
  2007) - Frontiers in Coding}. Lake Tahoe, California, USA.

\bibitem[\protect\citeauthoryear{Nepomniaschaya}{Nepomniaschaya}{2001}]{nepomn%
iaschaya:efficentEdmonds}
{\sc Nepomniaschaya, A.} 2001.
\newblock Efficient implementation of {E}dmonds' algorithm for finding optimum
  branchings on associative parallel processors.
\newblock In {\em Proc. of the Eighth Intern. Conf. on Parallel and Distributed
  Systems (ICPADS'01)}. KyongJu City, Korea.

\bibitem[\protect\citeauthoryear{Poor}{Poor}{1994}]{poor:detection+estimation-%
book}
{\sc Poor, H.~V.} 1994.
\newblock {\em {A}n {I}ntroduction to {S}ignal {D}etection and {E}stimation}.
\newblock Springer-Verlag.

\bibitem[\protect\citeauthoryear{Pradhan and Ramchandran}{Pradhan and
  Ramchandran}{1999}]{sandeep-kannan:discus}
{\sc Pradhan, S.~S.} {\sc and} {\sc Ramchandran, K.} 1999.
\newblock Distributed source coding using syndromes ({DISCUS}): Design and
  construction.
\newblock In {\em Proc. IEEE Data Compression Conf. (DCC)}. Snowbird, UT.

\bibitem[\protect\citeauthoryear{Pradhan and Ramchandran}{Pradhan and
  Ramchandran}{2005}]{Pradhan:CosetCodes}
{\sc Pradhan, S.~S.} {\sc and} {\sc Ramchandran, K.} 2005.
\newblock Generalized coset codes for distributed binning.
\newblock {\em IEEE Trans. Inform. Theory\/}~{\em 51}, 3457--3474.

\bibitem[\protect\citeauthoryear{Rebollo-Monedero, Zhang, and
  Girod}{Rebollo-Monedero
  et~al\mbox{.}}{2003}]{Rebollo-Monedero:QuantizersForDSC}
{\sc Rebollo-Monedero, D.}, {\sc Zhang, R.}, {\sc and} {\sc Girod, B.} 2003.
\newblock Design of optimal quantizers for distributed source coding.
\newblock In {\em Proceedings of the Data Compression Conference (DCC’03)}.

\bibitem[\protect\citeauthoryear{Scaglione and Servetto}{Scaglione and
  Servetto}{2002}]{ScaglioneS:02}
{\sc Scaglione, A.} {\sc and} {\sc Servetto, S.~D.} 2002.
\newblock On the interdependence of routing and data compression in multi-hop
  sensor networks.
\newblock In {\em Proc. ACM MobiCom}. Atlanta, GA.

\bibitem[\protect\citeauthoryear{Servetto}{Servetto}{2000}]{Servetto:02b}
{\sc Servetto, S.~D.} 2000.
\newblock Lattice quantization with side information.
\newblock In {\em Proc. IEEE Data Compression Conf. (DCC)}. Snowbird, UT.

\bibitem[\protect\citeauthoryear{Slepian and Wolf}{Slepian and
  Wolf}{1973}]{slepian-wolf:correlated-sources}
{\sc Slepian, D.} {\sc and} {\sc Wolf, J.~K.} 1973.
\newblock Noiseless coding of correlated information sources.
\newblock {\em IEEE Trans. Inform. Theory\/}~{\em IT-19,\/}~4, 471--480.

\bibitem[\protect\citeauthoryear{Strassen}{Strassen}{1969}]{strassen:gaussianE%
limination}
{\sc Strassen, V.} 1969.
\newblock Gaussian elimination is not optimal.
\newblock {\em Numer. Math.\/}~{\em 13}, 354--356.

\bibitem[\protect\citeauthoryear{Viswanathan and Berger}{Viswanathan and
  Berger}{1997}]{Berger:quadrGaussianCEO}
{\sc Viswanathan, H.} {\sc and} {\sc Berger, T.} 1997.
\newblock The quadratic {G}aussian {CEO} problem.
\newblock {\em IEEE Trans. Inform. Theory\/}~{\em 43}, 1549--1559.

\bibitem[\protect\citeauthoryear{Ward}{Ward}{1963}]{Ward}
{\sc Ward, J.} 1963.
\newblock Hierarchical grouping to optimize an objective function.
\newblock {\em Journal of the American Statistical Association\/}~{\em 58},
  236--244.

\bibitem[\protect\citeauthoryear{Xiong, Liveris, and Cheng}{Xiong
  et~al\mbox{.}}{2004}]{DSCforSensorNetw}
{\sc Xiong, Z.}, {\sc Liveris, A.~D.}, {\sc and} {\sc Cheng, S.} 2004.
\newblock Distributed source coding for sensor networks.
\newblock {\em IEEE Signal Processing Magazine\/}.

\bibitem[\protect\citeauthoryear{Zamir, Shamai, and Erez}{Zamir
  et~al\mbox{.}}{2002}]{zamir00nested}
{\sc Zamir, R.}, {\sc Shamai, S.}, {\sc and} {\sc Erez, U.} 2002.
\newblock Nested linear/lattice codes for structured multiterminal binning.
\newblock {\em Information Theory, IEEE Transactions on\/}~{\em 48,\/}~6,
  1250--1276.

\end{thebibliography}


\begin{thebibliography}{10}

\bibitem{AhoUllman:Foundations}
Alfred~V. Aho and Jeffrey~D. Ullman.
\newblock {\em Foundations of Computer Science, C Edition}.
\newblock W. H. Freeman \& Co., New York, NY, USA, 1997.

\bibitem{bar05}
J.~Barros and M.~Tuechler.
\newblock Scalable decoding on factor trees: A practical solution for sensor
  networks.
\newblock {\em IEEE Transactions on Communications}, 54:284--294, 02 2006.

\bibitem{Berger:CEO}
T.~Berger, Z.~Zhang, and H.~Viswanathan.
\newblock The {CEO} problem.
\newblock {\em IEEE Trans. Inform. Theory}, 42:887--902, 1996.

\bibitem{CardinalAssche:EntropyQuant}
Jean Cardinal and Gilles~Van Assche.
\newblock Joint entropy-constrained multiterminal quantization.
\newblock In {\em Proceedings of the International Symposium on Information
  Theory}, Lausanne, Switzerland, 2002.

\bibitem{Berger:CEORD}
Jun Chen, Xin Zhang, Toby Berger, and Stephen~B. Wicker.
\newblock An upper bound on the sum-rate distortion function and its
  corresponding rate allocation schemes for the {CEO} problem.
\newblock {\em Special Issue of JSAC, On Fundamental Performance of Wireless
  Sensor Networks}, May 2004.

\bibitem{ChuLiu:MDST}
Y.~J. Chu and T.~H. Liu.
\newblock On the shortest arborescence of a directed graph.
\newblock {\em Science Sinica}, 14:1396--1400, 1965.

\bibitem{cover-thomas:it-book}
T.~M. Cover and J.~Thomas.
\newblock {\em {Elements of Information Theory}}.
\newblock John Wiley and Sons, Inc., 1991.

\bibitem{dietrich97fast}
C.~R. Dietrich and G.~N. Newsam.
\newblock Fast and exact simulation of stationary {Gaussian} processes through
  circulant embedding of the covariance matrix.
\newblock {\em SIAM Journal on Scientific Computing}, 18(4):1088--1107, 1997.

\bibitem{Edmonds:MDST}
J.~Edmonds.
\newblock Optimum branchings.
\newblock {\em J. Research of the National Bureau of Standards}, 71B:233--240,
  1967.

\bibitem{FlynnG:87}
T.~J. Flynn and R.~M. Gray.
\newblock Encoding of correlated observations.
\newblock {\em IEEE Trans. Inform. Theory}, IT-33(6):773--787, 1987.

\bibitem{Georgiadis:MDST}
Leonidas Georgiadis.
\newblock Arborescence optimization problems solvable by {E}dmonds' algorithm.
\newblock {\em Theor. Comput. Sci.}, 301(1-3):427--437, 2003.

\bibitem{jain99:clustering}
A.~K. Jain, M.~N. Murty, and P.~J. Flynn.
\newblock Data clustering: a review.
\newblock {\em ACM Computing Surveys}, 31(3):264--323, 1999.

\bibitem{jayant-noll:coding-book}
N.~Jayant and P.~Noll.
\newblock {\em {D}igital {C}oding of {W}aveforms}.
\newblock Prentice Hall, 1984.

\bibitem{kschischang:factorgraphs}
F.~R. Kschischang, B.~Frey, and H.-A. Loeliger.
\newblock Factor graphs and the sum-product algorithm.
\newblock {\em IEEE Trans. Inform. Theory}, 47(2):498--519, 2001.

\bibitem{maierbacherB:CEOCoding}
G.~Maierbacher and J.~Barros.
\newblock Low-complexity coding for the {CEO} problem with many encoders.
\newblock In {\em Twenty-sixt Syposium on Information Theory in the Benelux},
  Brussels, Belgium, 2005.

\bibitem{maierbacherB:dio}
G.~Maierbacher and J.~Barros.
\newblock Diophantine index assignments for distributed source coding.
\newblock In {\em Proceedings of the 2007 IEEE Information Theory Workshop (ITW
  2007) - Frontiers in Coding}, Lake Tahoe, California, USA, 2007.

\bibitem{nepomniaschaya:efficentEdmonds}
Anna Nepomniaschaya.
\newblock Efficient implementation of {E}dmonds' algorithm for finding optimum
  branchings on associative parallel processors.
\newblock In {\em Proc. of the Eighth Intern. Conf. on Parallel and Distributed
  Systems (ICPADS'01)}, KyongJu City, Korea, 2001.

\bibitem{poor:detection+estimation-book}
H.~V. Poor.
\newblock {\em {A}n {I}ntroduction to {S}ignal {D}etection and {E}stimation}.
\newblock Springer-Verlag, 1994.

\bibitem{sandeep-kannan:discus}
S.~S. Pradhan and K.~Ramchandran.
\newblock Distributed source coding using syndromes ({DISCUS}): Design and
  construction.
\newblock In {\em Proc. IEEE Data Compression Conf. (DCC)}, Snowbird, UT, 1999.

\bibitem{Pradhan:CosetCodes}
S.~Sandeep Pradhan and Kannan Ramchandran.
\newblock Generalized coset codes for distributed binning.
\newblock {\em IEEE Trans. Inform. Theory}, 51:3457--3474, 2005.

\bibitem{Rebollo-Monedero:QuantizersForDSC}
David Rebollo-Monedero, Rui Zhang, and Bernd Girod.
\newblock Design of optimal quantizers for distributed source coding.
\newblock In {\em Proceedings of the Data Compression Conference (DCC’03)},
  2003.

\bibitem{ScaglioneS:02}
A.~Scaglione and S.~D. Servetto.
\newblock On the interdependence of routing and data compression in multi-hop
  sensor networks.
\newblock In {\em Proc. ACM MobiCom}, Atlanta, GA, 2002.

\bibitem{Servetto:02b}
S.~D. Servetto.
\newblock Lattice quantization with side information.
\newblock In {\em Proc. IEEE Data Compression Conf. (DCC)}, Snowbird, UT, 2000.

\bibitem{slepian-wolf:correlated-sources}
D.~Slepian and J.~K. Wolf.
\newblock Noiseless coding of correlated information sources.
\newblock {\em IEEE Trans. Inform. Theory}, IT-19(4):471--480, 1973.

\bibitem{strassen:gaussianElimination}
Volker Strassen.
\newblock Gaussian elimination is not optimal.
\newblock {\em Numer. Math.}, 13:354--356, 1969.

\bibitem{Berger:quadrGaussianCEO}
Harish Viswanathan and Toby Berger.
\newblock The quadratic {G}aussian {CEO} problem.
\newblock {\em IEEE Trans. Inform. Theory}, 43:1549--1559, 1997.

\bibitem{Ward}
J.~Ward.
\newblock Hierarchical grouping to optimize an objective function.
\newblock {\em Journal of the American Statistical Association}, 58:236--244,
  1963.

\bibitem{DSCforSensorNetw}
Z.~Xiong, A.~D. Liveris, and S.~Cheng.
\newblock Distributed source coding for sensor networks.
\newblock {\em IEEE Signal Processing Magazine}, 09 2004.

\bibitem{zamir00nested}
R.~Zamir, S.~Shamai, and U.~Erez.
\newblock Nested linear/lattice codes for structured multiterminal binning.
\newblock {\em Information Theory, IEEE Transactions on}, 48(6):1250--1276,
  2002.

\end{thebibliography}


\end{document}